\newif\ifdraft
\drafttrue
\ifdraft
\documentclass[]{sn-jnl}
\else
\documentclass[sn-aps]{sn-jnl}
\fi

\newlength\mywidth
\ifdraft
\else
\fi
\advance\mywidth by -1ex 

\usepackage{amsmath,amsfonts,amssymb}
\usepackage{algorithm}
\usepackage{array}
\usepackage{textcomp}
\usepackage{stfloats}
\usepackage{verbatim}
\usepackage{graphicx}
\usepackage{url}

\graphicspath{{./Images/}}
\usepackage[commentColor=mygreen]{algpseudocodex} 
\usepackage[caption=false,font=small,labelfont=rm,textfont=rm]{subfig}
\usepackage{caption} 

\usepackage{float}
\newfloat{algorithm}{tbp}{loa} 
\floatname{algorithm}{Algorithm}
\usepackage{color} 
\definecolor{mygreen}{rgb}{0.4,0.6,0.4}

\usepackage{hyperref} 
\ifdraft
	\usepackage{fancyhdr}
\fi

\def\Fig#1{{\bf Fig.~\ref{#1}}}
\def\Figu#1{{\bf Figure~\ref{#1}}}
\def\Tab#1{{\bf Tab.~\ref{#1}}}
\def\Tabl#1{{\bf Table~\ref{#1}}}

\def\Algo#1{{\bf Algorithm~\ref{#1}}}
\def\tdot{\!\cdot\!}
\def\D{\displaystyle}

\def\doublebox#1#2{%
  \hbox to \hsize{%
    \hfil
    \vtop{\hbox{#1}}%
    \hfil
    \vtop{\hbox{#2}}%
    \hfil
  }%
}
\renewcommand{\comment}[1]{}

\def\us{\underline{\phantom{x}}}

\begin{document}

\ifdraft
	\fontsize{11}{13}\selectfont 
	
\pagestyle{fancy}
\fancyhf{} 
\fancyhead[RO]{\today  \hfill Preprint\hfill \thepage}
\fancyhead[LE]{\thepage \hfill Strutz \& Rischke: Comparison of fast techniques for adaptive interval-based coding}

\fi

\title[]{Comparison of fast techniques for adaptive interval-based coding}

\author*[1]{\fnm{Tilo} \sur{Strutz}}\email{tilo.strutz]at[hs-coburg.de}

\author[2]{\fnm{Roman} \sur{Rischke}} 

\affil[1,2]{\orgname{Coburg University}, \orgaddress{\street{Friedrich-Streib-Str. 2}, \city{Coburg}, \postcode{96450}, \state{Germany}, \country{Bavaria}}
}





\abstract{
The transmission or storage of signals typically involves data compression. The final processing step in compression systems is generally an entropy coding stage, which converts symbols into a bit stream based on their probability distribution.
A distinct class of entropy coding methods operates not by mapping input symbols to discrete codewords (as in Huffman coding) but by operating on intervals or ranges. This approach enables a more accurate approximation of the source entropy, particularly for sources with highly skewed or varying symbol distributions. Representative techniques in this category include traditional arithmetic coding, range coding, and methods based on asymmetric numeral systems (ANS).
Due to their near-optimal compression efficiency and flexibility, these interval-based entropy coders are integral to modern data compression systems. 
The complexity of these methods depends mainly on three processing steps: the core routines of encoding and decoding doing the calculations, the interval-based determination of the correct symbol at decoder, and the efforts of keeping updated with respect to the varying symbol distribution.

The interval-based symbol determination at decoder typically demands for a searching procedure. In previous literature, it could be shown that the search can be replaced by a table-based approach with only ${\mathcal O}(1)$-complexity but having the side-effect that the adaptation of the multi-symbol statistic becomes infeasible because of the high time-consumption of adapting the table.

We compare different approaches to fast adaptive coding of multi-symbol alphabets and propose a new adaptation process using a special ring-buffer technique enabling the adaptive table-based decoding procedure in combination with the replacement of a division by a bit-shift operation at encoder and decoder core routines. 
The proposed technique is faster than alternative approaches for alphabets from about 12 to 64 different symbols when comparing the overall encoder+decoder time in a practical implementation.
} 


\maketitle

\ifdraft
\thispagestyle{fancy} \pagestyle{fancy}
\fi

\section{Introduction}
\label{sec_Introduction}
	%
General data compression, and video encoding in particular, can be very time-intensive when aiming for both high compression efficiency and excellent data quality. A significant portion of the processing time is often spent searching for optimal parameters. In video coding, this is most evident in the temporal prediction of image content, commonly referred to as motion estimation \cite{Jia19,Bos21,Liu22,Chou23,Gon23}. Another crucial step involves selecting the appropriate processing mode \cite{Sal20,Niu23,Sha23,Liu24}. Detailed analysis also shows that the entropy coding phase can involve computationally expensive procedures \cite{Pak20,Mer21} and efforts have been made to design efficient structures for hardware implementations \cite{Che15,Zhan19,Ram20,Cai22}.
	
Entropy coding is a core component of any data compression system. It operates by assigning a specific number or fraction of bits to each symbol $s_i$, based on its given or estimated probability $p(s_i)$. Ideally, the bit allocation reflects the symbols information content, defined as $I(s_i) = \log_2[1/p(s_i)]$.

\subsubsection*{Interval-based coding of multi-symbol alphabets}
Interval-based coding methods offer a practical approach towards this symbol-to-bit assignment. One of the earliest and most extensively studied techniques in this category is {\em arithmetic coding} (AC) \cite{Ris79}, which divides a numerical interval into subintervals proportional to the probabilities of the symbols. A closely related method is {\em range coding} (RC), introduced in the 1970s \cite{Mar79}. Although conceptually similar to arithmetic coding, range coding operates on digits from number systems with arbitrary bases, not limited to binary or decimal. In principle, arithmetic coding can be viewed as a special case of range coding. For practical reasons, range coding typically uses a base of $b\!=\!256$, enabling byte-wise input and output.
A third method in this category is based on asymmetric numeral systems (ANS) \cite{Dud14,Mof20}. Like RC, the range ANS (rANS) variant also processes data in bytes. However, it employs a first-in-last-out mechanism, meaning the encoder must handle data in reverse order compared to the decoder. This reversal introduces complexity in data management, particularly in adaptive coding scenarios \cite{Str23,She23}.

All three interval-based coding techniques are capable of handling multi-symbol alphabets with $K\ge 2$. This capability eliminates the need to convert symbols into binary sequences that require individual probability estimates -- an approach used in methods like context-based adaptive binary arithmetic coding (CABAC) \cite{Mar03,Sze12} and other binary coders \cite{Man05,Bel13,Aul16,Kim25} -- and can improve the throughput \cite{Han22}.

Efficient coding of large alphabets offers advantages in various applications and remains an active area of research \cite{Chen18,Str19,ChS20,Str23,Och24,Kri25}. This article focuses on coding schemes for alphabets with multiple symbols; binary approaches are not discussed further, since they do not belong to the addressed category of interval-based coding.

\begin{figure*}
\centering
\doublebox{\includegraphics[scale=0.15]{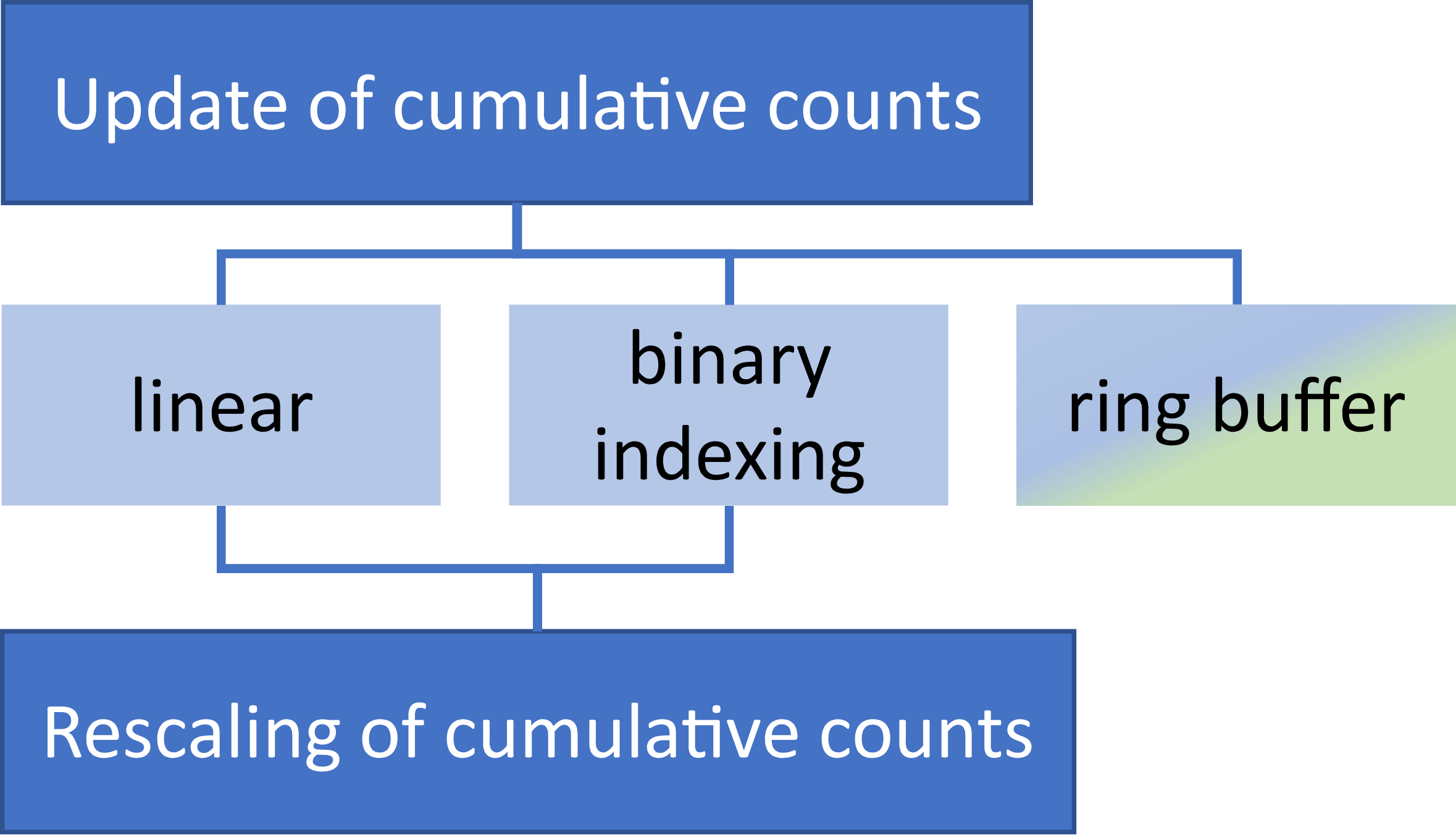}}
{\includegraphics[scale=0.15]{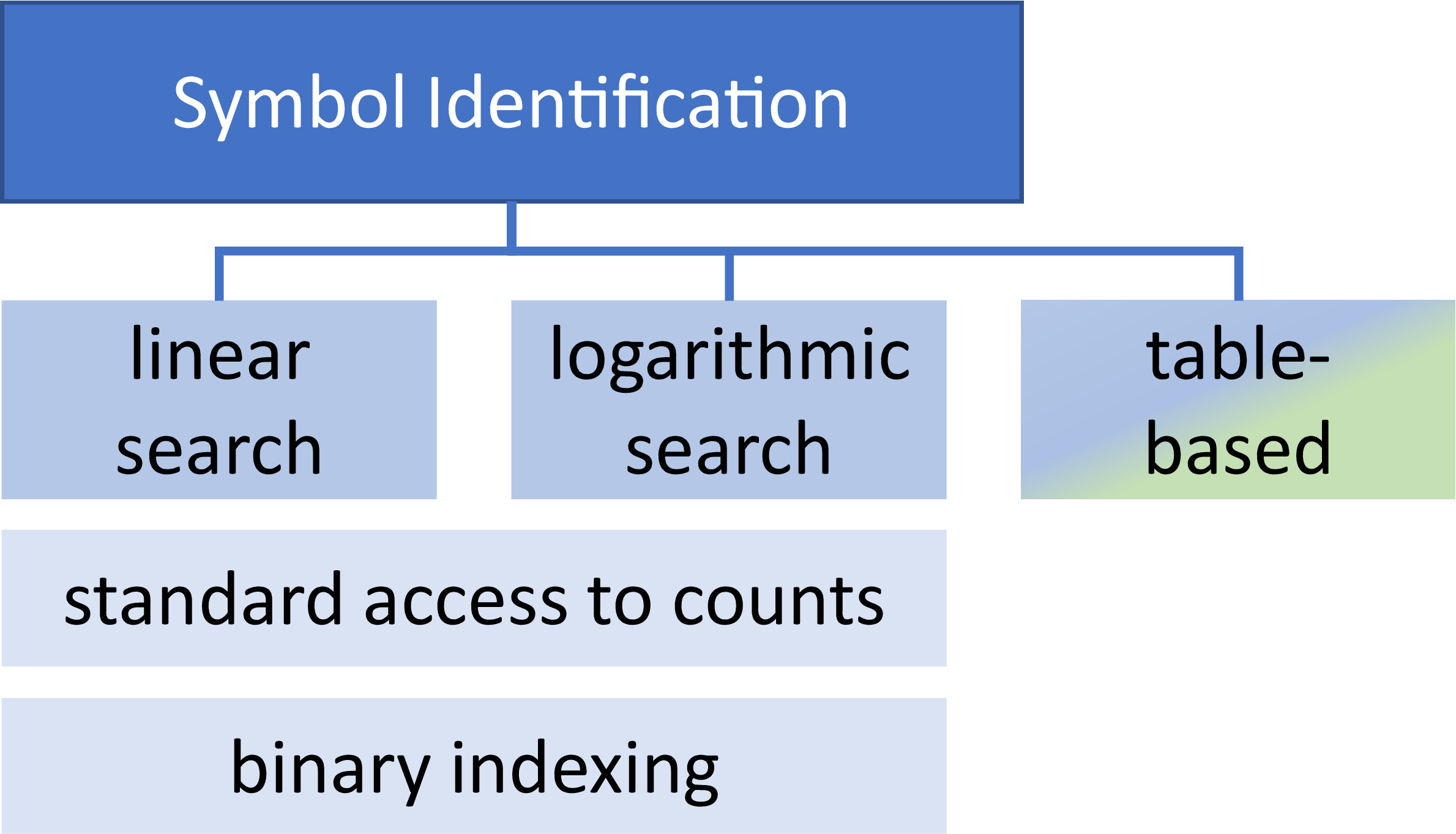}}

(a) \hfil\hfil (b)

	{\includegraphics[alt={This is the alternative text for screen readers.},scale=0.15]{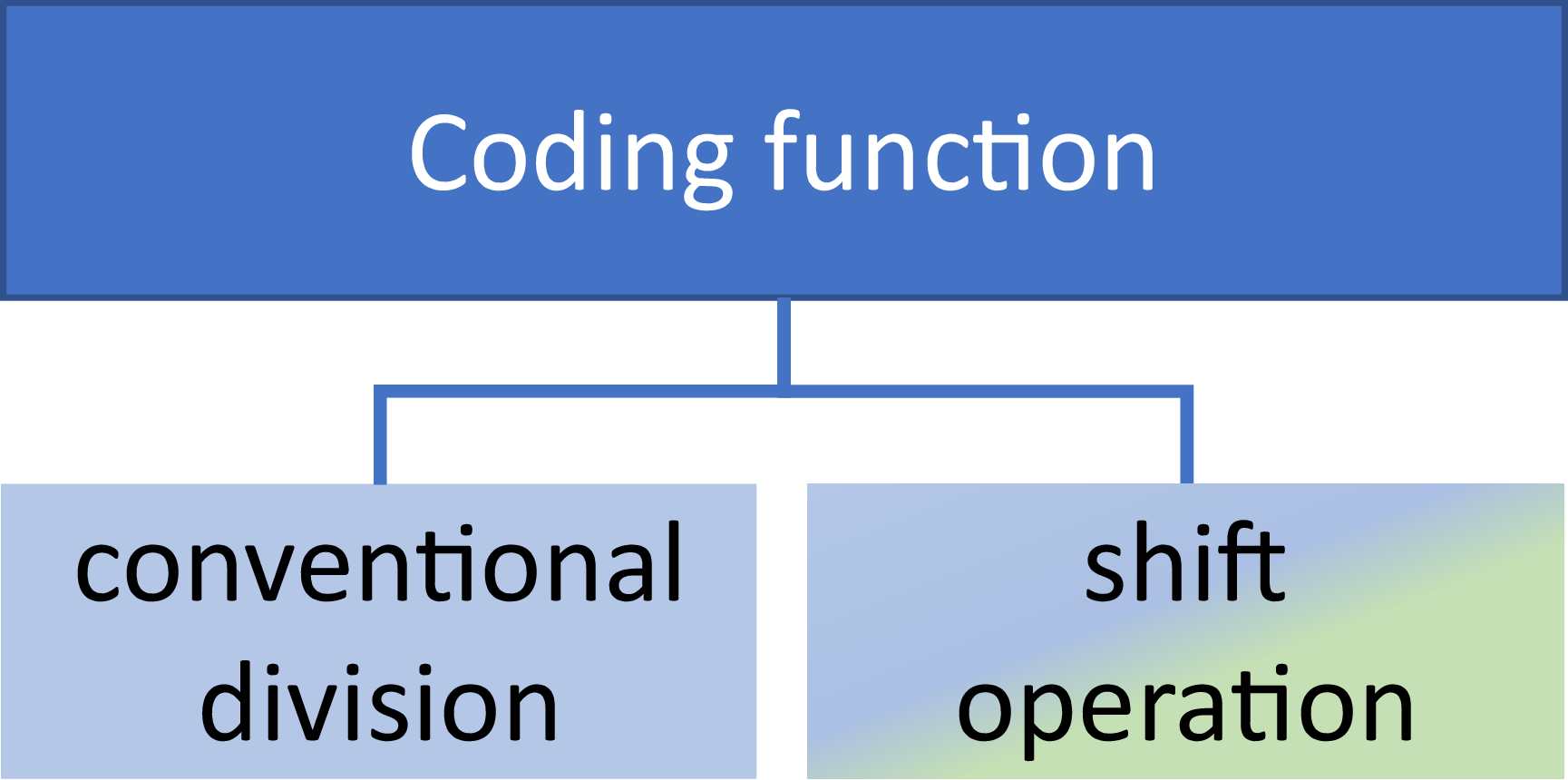}}

 (c)
 \caption{\label{fig_searchMethods}Methods considered for comparison: (a) methods for updating the symbol statistics, (b) search methods for symbol determination in decoder, (c) methods for symbol coding. Elements of the proposed method are marked in green.}
\end{figure*}

The multi-symbol methods have three components in common: (i) an adaptation process that creates the symbol distribution and keeps track of its changes, (ii) a procedure that finds the correct symbol regarding a computed internal status value at decoder, and (iii) the core calculations mapping the probability distribution to a range of numbers where each symbol corresponds to a subinterval.
\Figu{fig_searchMethods} summarizes the different approaches that are discussed in this paper.
	%

\subsubsection*{Updating the symbol statistics}
The symbol distribution has to be related to an array of cumulative counts. As the adaptive coding process typically starts with an assumed flat distribution, the array additionally has to be updated after processing each individual symbol \cite{Witten1987,Mof99}. Moreover, when the distribution changes within the symbol sequence, a mechanism should be implemented reducing the influence of old symbols and up-weighting newly occurring symbols, Figure \ref{fig_searchMethods}(a).
This can be achieved by regularly downscaling of counts \cite{Gallager1978,Schindler1998,Belyaev2017} or by using a limited number of past symbols \cite{Ghanbari1991,Ryabko1999}.

However, depending on the used alphabet size, updating the symbol statistics can significantly slow down the entire coding process if no special treatment is implemented \cite{Ryabko2003}. 
Some implementations forego an update step for each symbol and instead collect several symbols before any adjustment takes place \cite{Schindler1998}.
This is typically the case with ANS encoding, as its ‘last-in, first-out’ principle requires all symbols to be scanned before encoding can take place. Processing therefore usually takes place in blocks: for a block of a specific size, the symbols are counted, the distribution is transmitted to the decoder and the entire block is compressed in static mode (i.e. without adaptation within that block) \cite{Collet2015,Zstandard}. 
However, there are applications that do not allow for pre-scanning of the data and/or aim for maximum compression without having to transmit block-based overhead. In particular, when data exhibits strongly fluctuating statistical properties, fully adaptive processing can take these properties into account much more effectively and enable lower bit rates. For this reason, this paper focuses on fully adaptive coding.
	%
\subsubsection*{Symbol determination at decoder}
The problem of symbol identification at decoder, Figure \ref{fig_searchMethods}(b), typically involves searching routines which is a time-consuming issue for alphabets with $K\gg 2$. 
Logarithmic search has a lower $\mathcal{O}$-complexity than linear search, however in practical implementations, this advantage becomes active only if the number of intervals (i.e. the alphabet size) is high enough. 
Linear and logarithmic search can be combined with two different adaptation modes. The standard mode updates the statistic linearly ($\mathcal{O}(n)$) and the binary indexing (BI) mode, proposed by \cite{Fen94}, updates the statistic logarithmically ($\mathcal{O}(\log(n))$). 

The searching can only be avoided using a table that maps the internal status value directly to the corresponding symbol with $\mathcal{O}(1)$. This has been identified in \cite{Stru25} as best option in static compression mode where the symbols distribution remains fixed and the principle is already used in several coding systems as, for example, table-based ANS (tANS) \cite{Dud15}.

However, in adaptive processing of multi-symbol alphabets, not only the cumulative counts must be modified but also the mapping table, which makes this procedure too time-costly.%
\footnote{Note, all relevant binary arithmetic coders also operate table-based even in adaptive mode. Instead of updating a single table, several tables representing different distributions are prepared, which is not feasible for alphabets with an arbitrary number of symbols \cite{Howard1992}.}
Some approaches circumvent this problem by block-based static processing with symbol statistics separately determined and transmitted for each block.
From the point of view of compression efficiency, it would be preferable to omit this additional ancillary information and determine the symbol distribution ‘on the fly’, so that the estimated symbol statistics can quickly keep pace with the actual -- and potentially varying -- symbol distribution.
	%
\subsubsection*{Coding function}
The mapping of internal status information onto the array of cumulative counts used in interval-based coding methods typically requires at least one division operation, Figure \ref{fig_searchMethods}(c). The division can only be avoided if the divisor is a power of 2 enabling a bit-shift operation that accelerates the core of encoding and decoding \cite{Doshi2013,Hsi22}.

Unfortunately, this divisor is equal to the total count of all symbols when using conventional arithmetic or range coding. The arising problem is to keep this total count fixed while adapting the symbols distribution.
	%
\subsubsection*{Proposal}
	%
We propose a new, competitive method that enables fast, table-based symbol identification in the decoder, even in adaptive processing mode.
A specialised ring buffer technique enables the rapid processing of cumulative frequencies and updating of the decoder table, and allows the division to be replaced by a shift operation.
Instead of incrementing individual count values, the new approach works directly with the cumulative count values, thereby eliminating the time-consuming accumulation of count values.

\Tabl{tab_combinations} provides an overview of sensible combinations of methods and algorithms.
The effectiveness of the new approach is validated using simulated test data and real-world data in comparison with the other methods listed. 
\begin{table}[tb]
\centering
	\caption{\label{tab_combinations}Methods that can be combined in a meaningful manner}
\setlength{\tabcolsep}{3pt}	\begin{tabular}{|c||c|c|c|c|c|}
		\hline
				& Symbol 					& Shift			&Ring 		& 	 			& 		 			\\
	Mode	& Identification	& Operation	& Buffer	& BI 			& 		Label 			\\
	\hline                                                        
\hline                                                          
				&  linear search 	& on/off		& -				& no			&	fwd[Shift]		\\
static	&  log. search		& on/off		& -				& no 			&	log[Shift]		\\
				&  table-based		& on/off		& -				& no 			&	tab[Shift]			\\
\hline                                                          
				&  linear search 	& off				& yes/no	& yes/no	&	fwd[Ring][BI]	\\
				&  log. search 		& off				& yes/no	& yes/no 	&	log[Ring][BI]		\\
				&  table-based		& off				& yes/no	& yes/no 	&	tab[Ring][BI]			\\
adaptive&  linear search 	& on				& yes			& yes/no	&	fwdRingShift[BI]	\\
				&  log. search		& on				& yes			& yes/no 	&	logRingShift[BI]		\\
				&  table-based		& on				& yes			& yes/no 	&	tabRingShift[BI]			\\
\hline
	\end{tabular}
\end{table}
The  rightmost column contains the labels used for investigating the performance.
	%
\section{Basics}\label{sec_basics}
	%
The coding process operates over an interval $[0;M)$, which is partitioned into $K$ subintervals -- each corresponding to one of the $K$ distinct symbols $s_i$ in the alphabet, where $i\!=\!0,1,\dots,K\!-\!1$. 
The length of each subinterval is proportional to the frequency (the count) $h(s_i)$ or $h[i]$ of the respective symbol. 
These frequencies determine the boundaries of the subintervals, 
which are represented by the cumulative counts $h_{\rm k}[\cdot]$:
	\begin{align}\label{eq_sumCounts}
		h_{\rm k}[i+1] &= h_{\rm k}[i] + h[i] \quad\forall i=0,1,\dots,K-1\\
			&\quad\mbox{with}\quad h_{\rm k}[0]=0
			\quad\mbox{and}\quad  h_{\rm k}[K] = M = totalCount    \nonumber
	\;.
	\end{align}
In this context, $h_{\rm k}[i]$ represents the lower boundary, and $h_{\rm k}[i+1]$ the upper boundary of the subinterval assigned to symbol $s_i$. The symbols probabilities are estimated as relative frequencies $p(s_i)=p[i]=h[i] / h_{\rm k}[K]$.

On the decoder side, all three intervall-based coding methods share a common principle: a code value $c$ is inferred from the compressed data stream to identify the current subinterval, thereby determining the decoded symbol. To decode, $c$ is compared against the interval boundaries. If $c \in [h_{\rm k}[i], h_{\rm k}[i+1])$, then $s_i$ is the decoded symbol (see Section \ref{sec_symbolSearch}). 

In static coding, the symbol statistics are determined once on the encoder side and then transmitted to the decoder. Some symbol counts $h[i]$ may be zero, since the set of possible symbols is predefined and known in advance making it clear which symbols can or cannot occur. 
In adaptive coding, the process typically begins with a uniform distribution and $M\!=\!K$. Each time a symbol $s_i$ is encoded or decoded, its count $h[i]$ and automatically also $M$ are incremented.
Adaptive compression requires all symbol counts to be at least one, ensuring that every potential symbol can be distinguished. This means that the cumulative counts $h_{\rm k}[i]$ must differ, and no subinterval can have zero width.

According to Equation (\ref{eq_sumCounts}), when $h[i]$ is increased by one, all cumulative counts $h_{\rm k}[j]$ for $i<j \le K$ must also be updated. In the worst case, this means updating all $K$ cumulative values. For large alphabets, this update process can become a performance bottleneck, necessitating more efficient strategies. Unlike symbol search, this adaptation step affects both encoder and decoder, as they must maintain synchronized cumulative counts (see Section \ref{sec_updating}).

However, continuously updating the model by incrementing symbol counts can cause the total count $h_{\rm k}[K]\!=\!M$ to grow excessively. If $h_{\rm k}[K]$ becomes too large, it may interfere with the proper mapping of counts to the internal state variable used in the coding process. To prevent this, a maximum limit must be imposed on $h_{\rm k}[K]$ \cite{Witten1987}.
	%
\section{Symbol determination}\label{sec_symbolSearch}
	%
Identifying the correct symbol in the decoder requires a search procedure that compares a status variable with the boundaries of subintervals. 
This section explains three techniques that proved practical relevance in \cite{Stru25}. Figure \ref{fig_searchMethods}(b) gives an overview. 
	%
\subsection{Conventional searching methods}\label{sec_conventionalSearch}
	%
Linear search is the most straightforward approach for a decoder to locate the correct subinterval and determine the corresponding symbol. It works by sequentially comparing the code value $c$ with each interval boundary until the appropriate match is found.

When the search begins with the first symbol, it is known as a linear forward search,
 (see \Algo{alg_getSymbolLinearForward}, {\scshape getSymbolLinForward()}). 
    %
	    %
\begin{algorithm}
 \caption{\label{alg_getSymbolLinearForward}Determination of the correct symbol index $i$ using linear forward search or logarithmic search}
 \hfil
  \begin{minipage}[t]{\mywidth}
		\ifdraft
			\setlength{\baselineskip}{12pt} 
		\fi
		\begin{algorithmic}[1]
		\small
		\Function{getSymbolLinForward}{$K,c, h_{\rm k}[\cdot]$}
			\State $i \gets 1$
			\Comment{start search at first symbol $s_0$}
			\While {$c \ge h_{\rm k}[i]$}
				\Comment{if not inside the current sub-interval \dots}
				\State $i \gets i+1$
				\Comment{\dots go to next index}
			\EndWhile
			\State \Return $i-1$
		\EndFunction	
		\State{~}
		\Function{getSymbolLogarithmic}{$K,c, h_{\rm k}[\cdot]$}
			\State $bottom \gets 0; top \gets K$
			\Comment{outer interval boundaries}
			\Repeat 
				\State $i \gets (top + bottom) >> 1$
				\Comment{interval in the middle}
				\If {$c < h_{\rm k}[i]$}
					\State  $top \gets i$ \Comment{choose lower half}
				\Else
					\State $bottom \gets i+1$ 	\Comment{choose upper half}
				\EndIf
			\Until {$top == bottom$}
			\Comment{stop if no further subdivision is possible}
			\State \Return $bottom - 1$
		\EndFunction	
		\end{algorithmic}
	\end{minipage}
\end{algorithm}
	%
This method was introduced in a first practical implementation of source coding as described in \cite{Witten1987}. It is particularly efficient when symbols with lower indices have higher probabilities.

The worst-case complexity of $\mathcal{O}(K)$ can be reduced to $\mathcal{O}(\log{K})$ by a logarithmic search which is also known as bisection search, {\scshape getSymbolLogarithmic()}. 
	    %
	%
In this method, each comparison with an interval boundary effectively halves the number of remaining candidates.
	%
\subsection{Table-based searching}\label{subsec_tableBased}
	%
The need for an iterative search to locate a specific subinterval can be entirely eliminated by using a precomputed lookup table. This table directly maps each code value $c$ to its corresponding symbol index. To ensure full coverage, the array must have a size of at least the total symbol count $h_{\rm k}[K]$.

For instance, if there are four distinct symbols (i.e. $K=4$) with frequencies as shown in the following table
\begin{center}
\begin{tabular}{c|*4{c}}
  $i$		 	& 0 &	1	&	2	&	3	\\
	\hline
	$h[i]$	&	2	&	3	&	1	&	4
\end{tabular}\;,
\end{center}
the total count is equal to 10. The code value $c$ is an element of the half-open interval $[0,h_{\rm k}[K])$ and the lookup table $t[\cdot]$ would be
\begin{center}
\begin{tabular}{c|*{10}{c}}
  $c$							& 0 &	1	&	2	&	3	&	4	&	5	&	6	&	7	&	8	&	9	\\
	\hline
	$i=t[c]$							&	0	&	0	&	1	&	1	&	1	&	2	&	3	&	3	&	3	&	3	
\end{tabular}
\;.
\end{center}
Each symbol index $i$ appears exactly $h[i]$ times in the lookup table. 

This table-based method was found in \cite{Stru25} to be the best option in static compression mode, where the symbol distribution is determined once beforehand (see \Algo{alg_tableSearch}, procedure {\scshape  createTable()}) and remains unchanged throughout the process. 
\begin{algorithm}
 \caption{\label{alg_tableSearch}Required procedures of table-based search: initialization, mapping, updating}
 \hfil
  \begin{minipage}[t]{\mywidth}
		\ifdraft
			\setlength{\baselineskip}{12pt} 
		\fi
		\begin{algorithmic}[1]
		\small
		\Procedure{createTable}{$K, h[\cdot], t[\cdot]$}
			\State $idx \gets 0$, $i \gets 0$
			\Repeat 		\Comment {for all symbol indices}
				\State $cnt \gets 0$
				\Repeat  \Comment {write symbol index into table}
					\State $t[idx] \gets i$, $idx \gets idx +1$, $cnt \gets cnt + 1$
				\Until {($cnt == h[i]$)}\Comment {repeat $h[i]$ times}
				\State $i \gets i + 1$  \Comment {next symbol index}
			\Until {$i == K$}
			\State \Return $t[\cdot]$
		\EndProcedure	
		\State {~}
		\Function{getSymbolTableBased}{$c, t[\cdot]$}
			\State $i \gets t[c]$
			\State \Return $i$
		\EndFunction	
		\State {~}
		\Procedure{updateTable}{$K, t[\cdot], h_{\rm k}[\cdot], i$}
			\Repeat
				\State $idx = h_{\rm k}[i+1]$ \Comment {position to be changed}
				\State $t[idx] \gets i$	\Comment {set new symbol index $i$}
				\State $i \gets i + 1$  \Comment {go to next symbol index}
			\Until {$i == K$}
			\State \Return $t[\cdot]$
		\EndProcedure	
		\end{algorithmic}
	\end{minipage}
\end{algorithm}
The lookup operation has a constant time complexity of $\mathcal{O}(1)$, as implemented in Alg.~\ref{alg_tableSearch}, function {\scshape  getSymbolTableBased()}.
However, this efficiency comes at the cost of high memory usage, since the table must accommodate all possible code values. To manage this, it is advisable to keep the total symbol count $h_k[K]$ within a reasonable limit.

In adaptive mode, only $K$ entries need to be accessed in the worst case ($K/2$ on average for flat distributions) and the update complexity is $\mathcal{O}(K)$. Procedure {\scshape updateTable()} in Alg.~\ref{alg_tableSearch} shows the corresponding pseudo-code. 
	%
	%
\section{Core calculations in encoder and decoder}\label{sec_coder}
	%
In this paper, an implementation of a range coder is used for explaining and investigating the proposed approach.
\Algo{alg_RCcoding} shows the core coding routines.
\begin{algorithm}
 \caption{\label{alg_RCcoding}General encoder and decoder procedures of range coding}
  \hfil\begin{minipage}[t]{\mywidth}
	\begin{algorithmic}[1]
		\small
	\Procedure{RCencodeInit}{status variables}
		\State $low \gets 0$;  $buffer \gets 0$
		\State $countUndecidedDigits \gets 0$
		\State $range \gets (1 << (m\cdot 8) - 1$
				\Comment {number of digits ($m$) times bits per digit (8)}
		\State \Return status variables
	\EndProcedure	
	\State {~}
	\Procedure{RCencode}{status variables, $h_{\rm k}[\cdot]$, symbol $i$}
		\State $low \gets low + h_{\rm k}[i] \cdot (range /h_{\rm k}[K])$
			\Comment{update status variables}
		\State $range \gets (h_{\rm k}[i+1] - h_{\rm k}[i]) \cdot (range/h_{\rm k}[K])$
			\Comment{maintaining of $h[i]$ is not required}
		\While{($range < b^{m-1}$)}
			\Comment{as long as current range is too short,}
			\State \Comment{i.e. {\em range} comprises less than $\mbox{m}\!-\!1$ digits}
			\State \textsc{updateLow}(status variables)\Comment{scale {\em low} and output digits if necessary}
			\State $range \gets range\cdot b$
			\Comment{scale {\em range} by shifting one digit position}
		\EndWhile
		\State \Return status variables
	\EndProcedure	
	\State {~}
	\State {~}
	\Procedure{RCdecodeInit}{status variables}
		\State $range \gets (1 << (m\cdot 8) - 1$
				\Comment {number of digits ($m$) times bits per digit (8)}
		\State $code \gets 0$
		\For{($i\gets 0$; $i< m+1$; $i\gets i+1$)}
			\State{$code \gets code \cdot b + nextInputByte$}
			\Comment{read $m+1$ digits}
		\EndFor
		\State \Return status variables
	\EndProcedure	
	\State {~}
	\Procedure{RCdecode}{status variables, $h_{\rm k}[\cdot]$, symbol $i$}
		\State $c \gets code / (range /h_{\rm k}[K])$ 
			\Comment{map code to interval, $\frac{c}{h_{\rm k}[K]}\approx \frac{code}{range}$}
		\State $code \gets code - h_{\rm k}[i] \cdot (range / h_{\rm k}[K])$
		\State $range \gets (h_{\rm k}[i+1] - h_{\rm k}[i]) \cdot (range / h_{\rm k}[K])$ \Comment{narrow the interval}
		\While{($range < b^{m-1}$)}
			\Comment{as long current range is too short,}
			\State \Comment{i.e. {\em range} comprises less than $m\!-1\!$ digits}
			\State $code\gets code \cdot b + nextInputByte$	\Comment{scale and read next digit from stream}
			\State $range \gets range \cdot b $
		\EndWhile
		\State \Return status variables, c
	\EndProcedure	
	\end{algorithmic}
	\end{minipage}
\end{algorithm}
The procedure \textsc{RCencode}() requires the status variables ($low, range$), the distribution model $h_{\rm k}[\cdot]$, and the current symbol $s_i$ as input.
The variable $m$ is the number of digits that the internal status variable can comprise and $b$ is the basis of the number system used ($b=256$).
	
The cumulative probability is represented by
$p_{\rm k}[i] = h_{\rm k}[i] / h_{\rm k}[K]$. 
As in line 8 of procedure \textsc{RCencode}() can be seen, the integer division with the total count $h_{\rm k}[K]$ is performed first.
This slightly reduces the accuracy of computation, however in practical implementations, this has two benefits: (i) the quotient $range/h_{\rm k}[K]$ must then be computed only once and (ii) the integer multiplication does not run into overflows.
The procedure \textsc{updateLow}() is not of interest in this context and will not be discussed further.

At decoder, similar computations take place, Alg.\ \ref{alg_RCcoding}, {\scshape RCdecode()}.
	    %
	%
After the symbol has been determined (see Section \ref{sec_symbolSearch}), the decoder can modify the status variables in the same manner as the encoder.

The division $range / h_{\rm k}[K]$ is one of the most time-consuming operation at both encoder and decoder. It is obvious that the division can be replaced by a bit-shift operation $range >> \log_2( h_{\rm k}[K])$ if the total count $h_{\rm k}[K]$ is a power of two (i.e. $h_{\rm k}[K]=M=2^p$). However, this contradicts the idea of successive adaptation of symbol counts as explained in Section \ref{sec_basics}. Next section provides a solution to this problem.
	%

\section{Handling of cumulative counts}\label{sec_updating}
	%
In adaptive compression mode, the modification of cumulative counts is a matter of both the encoder and the decoder as already mentioned in Section \ref{sec_basics}.
We review conventional algorithms for updating the cumulative counts and present a new technique.
Afterwards we describe, how the cumulative counts are handled in static compression mode.
	%
\subsection{Linear update of cumulative counts}\label{subsec_linUpdate}
	%
When cumulative counts are stored sequentially in an array, updates can be performed using a single loop, as illustrated in \Algo{alg_updateSumCounts}. 
\begin{algorithm}
 \caption{\label{alg_updateSumCounts}Simple update of cumulative counts (with $\mathcal{O}(K)$ complexity) after processing of symbol $s_i$}
 \hfil
  \begin{minipage}[t]{\mywidth}
		\ifdraft
			\setlength{\baselineskip}{12pt} 
		\fi
		\small
		\begin{algorithmic}[1]
		\Procedure{updateSumCounts}{$K, h_{\rm k}[\cdot], i$}
			\If {$h_{\rm k}[K] \ge$ MAX\_TOTALCOUNT}
				\State {\scshape reScaleSumCounts($K, h_{\rm k}[\cdot]$)}
				\Comment{downscale values if needed}
			\EndIf
			\Repeat
				\State $i \gets i+1$
				\Comment{lower boundary of symbol must kept unchanged}
				\State $h_{\rm k}[i] \gets h_{\rm k}[i] + 1$
				\Comment{move upper boundary}
			\Until {$i == K$}\Comment{including last entry (totalCount)}
			\State \Return $h_{\rm k}[\cdot]$
		\EndProcedure	
		\end{algorithmic}
	\end{minipage}
\end{algorithm}
This algorithm is straightforward and relies on just a few basic operations. If symbols with higher indices tend to have higher probabilities, the update process requires only a few iterations and can be executed quickly.

However, when lower-indexed symbols are more frequent, the situation reverses, leading to slower performance. In the worst case, the time complexity is $\mathcal{O}(K)$, while for symmetric distributions, the average complexity is approximately $\mathcal{O}(K/2)$.
For large alphabets, this update mechanism may become a bottleneck in terms of processing speed, making it necessary to explore more sophisticated solutions.	
	%
\subsection{Updating of counts using binary indexing (BI)}
	%
A certain addressing technique can handle cumulative counts in a hierarchical manner and reduces the update complexity to $\mathcal{O}(\log(K))$. This comes with the cost of a more complex access to a single value $h_{\rm k}[i]$. 

Binary indexing, also known as binary index tree or Fenwick tree \cite{Fen94}, has been reviewed in detail in \cite{Stru25}. We would like to refer the reader to these articles.

When using the binary-indexing technique, the cumulative counts $h_{\rm k}[\cdot]$ cannot be accessed directly. Instead, dedicated functions are required. 
Following procedures are available: {\scshape getCountBI()}, {\scshape getSumCountBI()}, {\scshape getSymbolBI()}, {\scshape getSymbolLogarithmicBI()}, {\scshape updateSumCountsBI()}, and {\scshape reScaleSumCountsBI()}.
It is important to understand that only {\scshape updateSumCountsBI()} speeds up processing, since it reduces the number of accesses from $\mathcal{O}(K)$ to $\mathcal{O}(\log K)$. All other functions increase the computational load, because the cumulative counts $h_{\rm k}[i]$ must be calculated as sums based on the binary index tree. 
	%
\subsection{Proposed ring-buffer scheme for adaptive coding mode}\label{subsec_ringbuffer}
	%
The proposed method uses a ring-buffer (or sliding window) scheme: when the count of a new symbol is incremented then the count of the oldest symbol has to be decremented. 
Ryabko proposed in \cite{Ryabko1999} to decrement the count of a randomly chosen symbol in order to avoid extra memory for the buffer. Unfortunately, this is not practical as the count of the selected symbol could have already reached its minimum value ($=1$) making the search for another symbol necessary and contradicting the objective of fast processing.
Other existing methods initialise the symbol counts such that their distribution is almost flat and the total count $h_k[K]$ is equal to $2^p$ supporting shift operations instead of divisions \cite{Schindler1998,Belyaev2017} as had been explained in Section \ref{sec_coder}. 
This has the disadvantage that all symbols start with a relatively high count and the convergence towards possible skewed distributions is rather slow.  

We propose to start with an empty ring buffer. The processed symbols are successively stored until the total count $h_{\rm k}[K]$ reaches a desired value $M=2^p$. 

Since the coding process must start with an initialised array of cumulative counts, i.e. $h_{\rm k}[j]\! >\! h_{\rm k}[i]~ \forall j\!>\!i\!\ge\! 0$, the required buffer size is only $M\!-\!h_{\rm k}[K]$. Typically the process starts with counts $h[i]\!=\!1 ~\forall i$ leading to $h_{\rm k}[K]\! =\!K$. Therefore, after $M\!-\!K$ symbols have been processed, the buffer should be completely filled and the total count $h_{\rm k}[K]$ is a power of two. Now, the coding procedures (Alg. \ref{alg_RCcoding}) can switch from the conventional division operation to the bit-shift operation.

\Algo{alg_RCinitAdaptiveModel} depicts the initialization of the ring buffer and the mapping table.
\begin{algorithm}
 \caption{\label{alg_RCinitAdaptiveModel}Proposed initialization 
for adaptive processing}
  \hfil\begin{minipage}[t]{0.99\hsize}
	\begin{algorithmic}[1]
		\small
	\Procedure{RCinitAdaptiveModel}{$h_{\rm k}[\cdot]$, $t[\cdot]$, $K$, total count $M$}
		\State $h_{\rm k}[0] \gets 0$ \Comment{lower border of entire interval}
		\For{($i \gets 0$; $i < K$; $i\gets i+1$)}
			\State $h_{\rm k}[i+1] \gets h_{\rm k}[i] + 1$
			\Comment{subintervals have width equal to 1}
		\EndFor
		\State $ringBufLen \gets M-K$ \Comment{length of ring buffer}
		\State $ringBuf\!fer \gets $ allocateMemory($ringBufLen$)
		\For{($i \gets 0$; $i < ringBufLen$; $i\gets i+1$)}
			\State $ringBuf\!fer[i] \gets K$
			 \Comment{fill ring buffer with a non-symbol value}
		\EndFor
		\State $ringIdx \gets 0$ \Comment{start with first element of ring buffer}
		\LComment{for decoder only: ---------------------------}
		\State $t[\cdot] \gets $ allocateMemory($M$)
		\State $idx \gets 0$
		 \Comment{start position in table}
		\For{($i \gets 0$; $i < K$; $i\gets i+1$)}
			\State $count \gets h_{\rm k}[i+1] - h_{\rm k}[i]$\Comment{derive count of $s_i$}
			\For{($j \gets 0$; $j < count$; $j\gets j+1$)}
				\State $t[idx+j] \gets i$
				\Comment{fill table with current symbol}
			\EndFor
				\State $idx \gets idx + count$
				\Comment{go to region of next symbol}
		\EndFor
		\State \Return $t[\cdot]$, $h_{\rm k}[\cdot]$, $ringBuf\!fer[\cdot]$, $ringBufLen$
	\EndProcedure	
\end{algorithmic}
\end{minipage}
\end{algorithm}
During the coding process, both arrays, $ringbuf\!fer[\cdot]$ and $t[\cdot]$, have to be updated, \Algo{alg_RCupdateRingModel}.
\begin{algorithm}
 \caption{\label{alg_RCupdateRingModel}Proposed update of modelling structures for adaptive processing}
  \hfil\begin{minipage}[t]{0.99\hsize}
	\begin{algorithmic}[1]
		\small
	\Procedure{RCupdateModel}{$h_{\rm k}[\cdot]$, $t[\cdot]$, $K$, symbol $i$}
		\State $ oldSym\gets ringbuf\!fer[ringIdx]$ \Comment{get symbol to be removed}
		\State $ringbuf\!fer[ringIdx] \gets i$ \Comment{overwrite with new}
		\State $ringIdx \gets ringIdx + 1$ \Comment{go to position of next oldest symbol}
		\If {$ringIdx == ringBufLen$}
			\State $ringIdx \gets 0$ \Comment{end of array is reached, go to first position }
		\EndIf
		\If {$oldSym < i$}
			\For{($j \gets oldSym+1$; $j \le i$; $j\gets j+1$)}
				\State $h_{\rm k}[j] \gets h_{\rm k}[j] - 1$
				\Comment{move interval boundaries downwards}
				\State $t[h_{\rm k}[j]] \gets j$ \Comment{only at decoder: table update}
			\EndFor
		\Else
			\For{($j \gets i+1$; $j \le oldSym$; $j\gets j+1$)}
				\State $t[h_{\rm k}[j]] \gets j-1$ \Comment{only at decoder: table update}
				\State $h_{\rm k}[j] \gets h_{\rm k}[j] + 1$
				\Comment{move interval boundaries upwards}
			\EndFor
		\EndIf
		\State \Return $t[\cdot]$, $h_{\rm k}[\cdot]$, $ringBuf\!fer[\cdot]$
	\EndProcedure	
\end{algorithmic}
\end{minipage}
\end{algorithm}
The oldest symbol is extracted from the ring buffer as it influences the modification of the cumulative counts. Then it is overwritten by the current symbol and the ring-buffer pointer $ringIdx$ is moved to the next position. Now, the count of the removed symbol has to be decremented and the count of the new symbol has to be incremented. 
And this is exactly where the innovation enters the picture.
The  required modifications of the array of cumulative counts depend on the order of these two symbols. 
If the old one has a lower index, its upper bound must be decremented (i.e. its count gets smaller) and the lower bound of the new symbol must be decremented (its count gets higher). All values $h_{\rm k}[j]$ in between are also decremented in order to keep all other interval widths the same.
In contrast, if the new symbol $i$ has the lower index, all lower bounds from $i+1$ up to the old symbol are incremented, leading to an increased count of the new and a decreased count of the old symbol, while keeping all other counts untouched. In ideal case, the new symbol is equal to the old one and nothing has to be changed.
A main advantage of this procedure is that cumulative counts and the decoder-table entries are commonly modified in a loop reducing significantly the efforts of adapting the distribution. An extra array of counts $h[i]$ is not required.

It can be shown that for flat distributions the average number of modifications is only $K/3$ instead of $K/2$ when using Algorithm \ref{alg_RCupdateRingModel}, see Appendix \ref{app_uniform}. This could be verified by counting the accesses during the coding process.
	%
If the distribution is skewed, the computational load is even lower, regardless to which side it is skewed. The geometric distribution used in the investigations (Section \ref{sec_investigations}) demands only about $0.09\cdot K$ accesses on average while the standard approach requires $0.09\cdot K$ or $0.91\cdot K$ depending to which side the distribution is skewed, see Appendix \ref{app_geometric}.
%
	%
Note that updating the decoder mapping table can be handled simultaneously with updating the cumulative counts (Alg. \ref{alg_RCupdateRingModel}, lines 10 and 13) and requires less efforts than procedure {\scshape updateTable()} in Alg. \ref{alg_tableSearch}.

We focus on sources with varying symbol distributions, where adaptive coding is essential for low bitrates.
It is hardly possible to derive theoretical limits for compression effectiveness, as these always depend on the rate of change in the statistical distributions. It is obvious that, for memoryless sources, a symbol buffer whose size is at least equal to the number of symbols to be processed would be optimal for best compression. The faster the symbol distribution changes within a sequence, the fewer symbols the buffer should contain, in order to allow rapid adaptation to the current symbol probabilities.

	%
	
	%
\subsection{Proposed procedure in static coding mode}\label{subsec_staticScaling}
	%
In static processing mode, we neither have to care about updating the array of cumulative counts nor the mapping table.
The only point to be discussed here for improving the efficiency of the range coder is to replace the division with a shift operation, as already suggested in Section \ref{sec_coder}.

To achieve this, the symbol distribution must first be derived from the input sequence by counting the occurrences of each symbol. Afterwards, the counts need to be scaled so that the total count matches the desired target value $M=2^p$, \Algo{alg_RCscaleModel}.
\begin{algorithm}
 \caption{\label{alg_RCscaleModel}Proposed scaling of the symbol distribution to the desired total count in static coding mode}
  \hfil\begin{minipage}[t]{0.99\hsize}
	\begin{algorithmic}[1]
		\small
	\Procedure{RCscaleModel}{$h[\cdot]$, $h_{\rm k}[\cdot]$, $K$, $totalCount$, desired total count $M=2^p$}
		\State $sum \gets 0$ \Comment{current total count}
		\For{($i \gets 0$; $i < K$; $i\gets i+1$)}
			\Comment{determine number of non-zero counts}
			\State $cnt \gets 0$
			\If {$h[i] > 0$}
				\State{$cnt \gets (h[i]\cdot M + (totalCount >> 1)) / totalCount$}
				\If {$cnt == 0$} {$cnt \gets 1$} \Comment{must be at least equal to one}
				\Else {~~$h[i] \gets 1$} \Comment{if count was non-zero, then keep it non-zero}
				\EndIf
			\EndIf
			\State $sum\gets sum + cnt$ 
			\State $h_{\rm k}[i+1] \gets cnt$ \Comment{save intermediate counts, sum up later} 
		\EndFor
		\While {$sum > M$}
			\State{$best\us i \gets K$}
				\Comment{look for a symbol where decrementing count is harmless}
			\For{($i \gets 0$; $i < K$; $i\gets i+1$)}
				\State{$q \gets h_{\rm k}[i+1]$} \Comment{get intermediate count}
				\If {$q>=2$} 
						\Comment{only if there is something do decrement}
					\If {$best\us i == K ~||~$ \Comment{take first possible or \dots}\\
							$\phantom{if~} h[i] \cdot (2\cdot h_{\rm k}[best\us i+1]-1) < 
							h[best\us i]\cdot (2\cdot q-1)$} 
						\State $best\us i \gets i$
					\EndIf
				\EndIf
			\EndFor
			\State $h_{\rm k}[best\us i +1]\gets h_{\rm k}[best\us i+1] - 1$
			\State $sum\gets sum - 1$
		\EndWhile
		\While {$sum < M$}
			\State{$best\us i \gets K$}
				\Comment{look for a symbol where incrementing count is harmless}
			\For{($i \gets 0$; $i < K$; $i\gets i+1$)}
				\State{$q \gets h_{\rm k}[i+1]$} \Comment{get intermediate count}
				\If {$q>0$} 
						\Comment{$q==1$ is be fine for increment }
					\If {$best\us i == K ~||~$ \Comment{take first possible or \dots} \\
							$\phantom{if~} h[i] \cdot (2\cdot h_{\rm k}[best\us i+1]+1) < 
							h[best\us i]\cdot (2\cdot q+1)$} 
						\State $best\us i \gets i$
					\EndIf
				\EndIf
			\EndFor
			\State $h_{\rm k}[best\us i+1]\gets h_{\rm k}[best\us i+1] + 1$
			\State $sum\gets sum + 1$
		\EndWhile
		\State{$h_{\rm k}[0] \gets 0$} \Comment{compute cumulative counts}
		\For{($i \gets 0$; $i < K$; $i\gets i+1$)}
			\State {$h[i]\gets h_{\rm k}[i+1]$} \Comment{get intermediate count, needs to be transmitted}
			\State {$h_{\rm k}[i+1]\gets h_{\rm k}[i] + h[i]$}
					\Comment{accumulate counts}
		\EndFor
		\State \Return $h_{\rm k}[\cdot]$, $h[\cdot]$
	\EndProcedure	
\end{algorithmic}
\end{minipage}
\end{algorithm}
After the simple scaling in the first for loop, the resulting sum of the count values must be compared with $M$. The algorithm searches for the count values to be modified that result in the least loss of compression effectiveness. If this one-off search proves too time-consuming for a particular application, it would also be possible to increase or decrease any counter value, which would result in a marginally increased bit rate.
	%
\comment{
\begin{algorithm}
 \caption{\label{alg_RCscaleModel}Proposed scaling of the symbol distribution to the desired total count in static coding mode}
  \hfil\begin{minipage}[t]{0.99\hsize}
	\begin{algorithmic}[1]
		\small
	\Procedure{RCscaleModel}{$h[\cdot]$, $K$, $totalCount$, desired total count $M=2^p$}
		\State $n \gets 0$ \Comment{number of used symbols}
		\For{($i \gets 0$; $i < K$; $i\gets i+1$)}
			\Comment{determine number of non-zero counts}
			\If {$count == 0$}
				\State $n\gets n + 1$ 
			\EndIf
		\EndFor
		\State $R \gets M - n$ \Comment{ensures $totalCount == M$}
		\State $usedSymbols \gets 0$ , $sum \gets 0$ \Comment{}
		\State $h_{\rm k}[0] \gets 0$ \Comment{lower border of first sub-interval}
		\For{($i \gets 0$; $i < K$; $i\gets i+1$)}
				\Comment{compute all cumulative counts}
			\State $sum \gets sum + h[i])$
			\If {$h[i] > 0$} \Comment{if original count was non-zero ...}
					\State $usedSymbols \gets usedSymbols + 1$ \Comment{... than increment}
			\EndIf
			\State $h_{\rm k}[i+1] \gets usedSymbols + (sum \cdot R + (totalCount >> 1)) / totalCount $
			\State $h[i] \gets h_{\rm k}[i+1] - h_{\rm k}[i]$
			\Comment{re-compute counts}
		\EndFor
		\State \Return $h_{\rm k}[\cdot]$, $h[\cdot]$
	\EndProcedure	
\end{algorithmic}
\end{minipage}
\end{algorithm}
} 
	

The counts must be transmitted to the decoder and then both, encoder and decoder, are able to use the shift-operation-based coding routines.

The static coding mode is required  in the investigation section to quantify the influence of single  components of the proposed scheme.

	%
\section{Investigations and Results}\label{sec_investigations}
	%
All investigations were conducted using a range coder implemented in ANSI-C (base $b=256$) that is integrated into the algorithmic approaches discussed.
The precision of the coding engine is set to six digits, i.e. 48 bits in total.	
The modelling of data distribution uses a total count of $M=2^p$ with $p=12$ per default, which as been proven as good basis for modelling the statistics of typical data in image compression systems. 
This means, if the alphabet size is $K=32$, for example, the maximum count per symbol would be $2^{12-5}=128$ for a uniform distribution.
For particular investigations, $p$ will also be varied.

		%
\subsection{Methodology}
	%
The programs were executed on a notebook equipped with an Intel i7-1165G7@2.80GHz processor. To measure execution times, we used the \verb+__rdtsc()+ function, which returns the processor’s time stamp counter representing the number of clock cycles since the last system reset \cite{Mic24}.
Two time stamps were recorded: one immediately before the data encoding begins (including the setup of all necessary arrays for modelling), and another right after all symbols have been processed. The difference between these two values is taken as the time measurement.
It is important to note that all input and output data were handled in memory. Consequently, file I/O operations such as reading from or writing to disk were excluded from the timing measurements.
Measuring the time offers a practical comparison, taking all influences of computer systems into account, which theoretical considerations of the complexity cannot.

The measured execution time does not necessarily correspond to the actual number of CPU instructions executed during the coding process. Two key factors must be considered. First, the CPU may be concurrently handling other background tasks, which can affect timing accuracy. Second, the execution time is influenced by the CPU’s clock frequency, which is not constant.
Modern processors can dynamically adjust their clock speed boosting performance during short, intensive workloads. However, under sustained high loads, such as in our simulations, the CPU temperature may rise significantly, triggering thermal throttling. This mechanism reduces the clock frequency to prevent overheating, and it remains in effect until the temperature drops below a safe threshold.
As a result, the CPU frequency can fluctuate frequently, making time-based measurements unreliable indicators of actual computational effort.

To minimize the impact of background processes, all unnecessary applications were closed, the internet connection was disabled, and each measurement was repeated five times, with the lowest value selected for analysis. The issue of fluctuating CPU frequency was addressed by limiting the maximum processor performance to 75\% via the Windows 11 power settings. This adjustment resulted in a nearly constant CPU frequency of approximately 2.0 GHz, as confirmed through the task manager.
	
The experiments use ten data sequences, each comprising $10^8$ symbols \cite{Dat25}. The sequences differ in the number of the alphabet size, which is chosen as $K=2^n$ with $n=1,2,\dots, 10$.
As the algorithm performance usually depends on the symbol distribution, two typical variants are taken into account: uniform (flat) distribution with $p(s_i) = 1/K~\forall i$ and truncated geometric distribution with
		\begin{align}\label{eq_geometricDistribution}
			p(s_i) = \frac{(1-p)\tdot p^i}{1-p^K} \,,\; p = 2^{-1/2^k}, \;
			i=0, 1, \dots, K\!-\!1
			\,.
		\end{align}
This type of distribution has already been used by others, e.g. \cite{Sai04,Stru25}, and is representative for a large class of data.
The generating process is shown in \Algo{alg_generateGeometric}.
	%
\begin{algorithm}
 \caption{\label{alg_generateGeometric}Procedure of generating a sequence with geometrically distributed symbols}
 
  \hfil\begin{minipage}[t]{\mywidth}
	\begin{algorithmic}[1]
		\footnotesize
	\Procedure{GenGeometric}{length of sequence $N$, alphabet size $K$, parameter $k$}
		\State $p \gets 2^{-1/2^k}$ \Comment {define probability based on $k$}
		\For {$i = 0: K-1$}\Comment {compute single probabilities}
			\State $pr[i] \gets (1-p) \cdot p^i /(1-p^K)$
		\EndFor
		\State $prSum[0] \gets 0$\Comment {compute cumulative probabilities}
		\For {$i = 0: K-1$}
			\State $prSum[i+1] \gets prSum[i] + pr[i]$
		\EndFor
		\State $arr[\cdot] \gets$ array with $N$ uniformly distributed values
		\State $\qquad \qquad \in [0;1)$
		\For {$n = 0: N-1$} \Comment {for all sequence elements}
			\State {$i \gets 0$}
			\While{$arr[n] \ge prSum[i+1]$}
			  
				\State {$i \gets i + 1$}\Comment {search corresponding symbol}
			\EndWhile
			\State $sequence[n] \gets i$
		\EndFor
		\State \Return $sequence[\cdot]$
	\EndProcedure	
	\end{algorithmic}
	\end{minipage}
\end{algorithm}
	%

The parameter $k$ was selected such that the probability of symbols decreases significantly with increasing $i$ and the symbol counts remain above zero:
		\begin{align}\label{eq_kbasedOnK}
			k(K) = \max(0, \lfloor\log_2(K)\rfloor - 4)
			\;.
		\end{align}
		%
	%
	%
	%
\subsection{Static processing mode}
	%
Although the target application of the proposed scheme is adaptive coding, investigations on static coding are helpful revealing the influence of single components of this scheme.	
The static mode determines the symbol distribution in a preprocessing step by counting all symbols once. This statistic remains fixed for all symbols to be processed.
Thus on encoder side, only the influence of the shift operations on the processing speed is measured. On decoder side, also the algorithm for symbol identification has an impact.
	%
\subsubsection{Uniform distribution}
\begin{figure}
		\centering
	\subfloat[]{\includegraphics[scale = 1.]{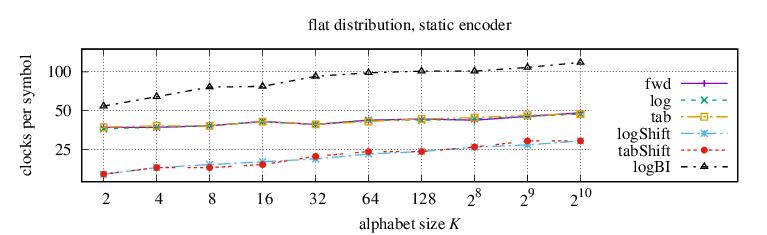}}\\[-0.3em]

  \subfloat[]{\includegraphics[scale = 1.]{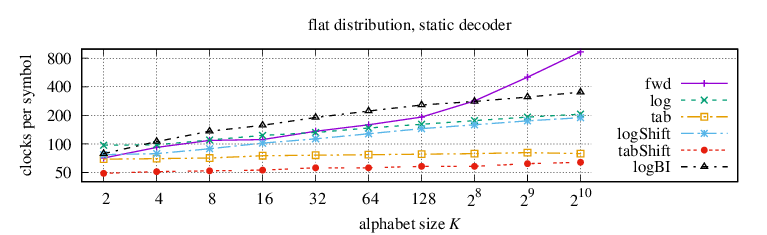}}
	\caption{\label{fig_clocksPerSymbolStaticFlat}Required processor clock cycles per symbol dependent on the alphabet size: flat distribution, $h(s_i) = h(s_j) \forall (i,j)$: (a) encoder, (b) decoder}
\end{figure}

	%
\Figu{fig_clocksPerSymbolStaticFlat} presents the number of clocks per symbol determined in dependence on the symbol alphabet when coding symbols with equal probabilities.
The labels are chosen according Table \ref{tab_combinations}.
Since `fwd', `log', and `tab' discriminate between different algorithms for symbol determination at decoder, they have no influence on the encoder and the curves are in principle identical in Figure \ref{fig_clocksPerSymbolStaticFlat}(a). 
The ring buffer is required for adaptation only, therefore it does not appear here.
The only discrimination in this investigation tangents the substitution of a division by a bit-shift operation.
It saves about 20 clocks per symbol corresponding to a reduction by 40\% - 50\% at the encoder.
Additionally, the binary indexing scheme `logBI' has been investigated. It performs the worst, as the hierarchical structure used to store the cumulative counts results in additional overhead when accessing the values, without offering any benefit.
		
The curves in Figure \ref{fig_clocksPerSymbolStaticFlat}(b), illustrating the behaviour of the decoder, show a little more variation, as symbol determination also influences the processing time.
Table-based symbol determination has almost constant processing times regardless the alphabet size because its complexity is $\mathcal{O}(1)$.
The switch from integer division to shift operations has the same positive effect as for the encoder.
	%
\subsubsection{Geometric distribution}
\begin{figure}\centering
	\subfloat[]{\includegraphics[scale = 1.]{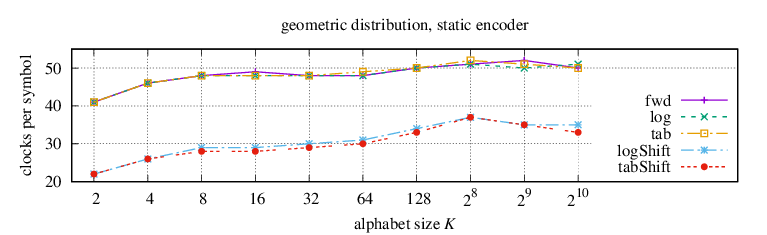}}\\[-0.3em]

	\subfloat[]{\includegraphics[scale = 1.]{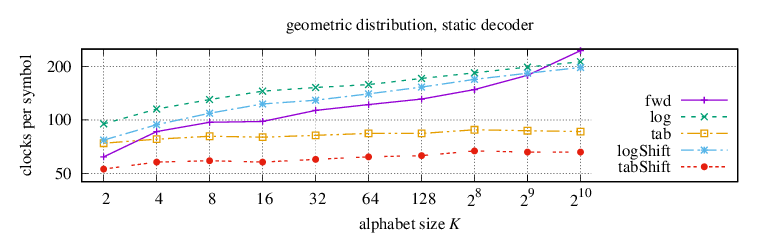}}
\caption{\label{fig_clocksPerSymbolStaticGeometric}Required processor clock cycles per symbol dependent on the alphabet size $K$ for geometric symbol distributions: (a) encoder, (b) decoder}
\end{figure}

	%
The results of the timing measurements for the compression of data with a geometric distribution are shown in \Figu{fig_clocksPerSymbolStaticGeometric}.
The encoder curves are about the same compared to the experiments with uniform distributions. 
At decoder side, the linear forward search can benefit from the distribution that is skewed towards symbols with small indices.
The method `tabShift' is the clear winner when operating in static mode as expected.
   	%
\subsection{Adaptive processing mode}
	%
Using the same data sets, the investigation were repeated, running the encoding/decoding software in adaptive mode. 
The coding times now additionally depend on the adaptation method.

In conventional adaptation mode, the software starts with a uniform distribution and, after processing each symbol, its count is incremented. When a total count of $2^{12}$ is reached, all counts are downscaled by factor 2.
The proposed adaptation mode utilizes a ring buffer of size $2^{12}-K$. As long as the buffer is not yet completely filled, the behaviour is like in the conventional mode. Then, the total count remains fixed, since the increment of a new symbol's count is combined with the decrement of the oldest symbol's count. This is also the moment were the coding routines replace the division by the shift operation. As the number of symbols in the test data ($10^8$) is much higher than the buffer legth, the delayed switch to shift operations has almost no influence on the time measurement. 

The second option is using the binary indexing that allows updating the cumulative counts with $\mathcal{O}(\log K)$ instead of  $\mathcal{O}(K)$. 
	%
\subsubsection{Uniform distribution}
\begin{figure}
	\centering
	\subfloat[]{\includegraphics[scale = 1.]{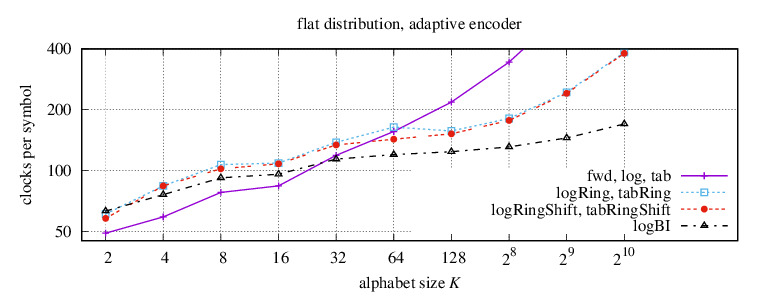}}\\[-0.3em]
	
	\subfloat[]{\includegraphics[scale = 1.]{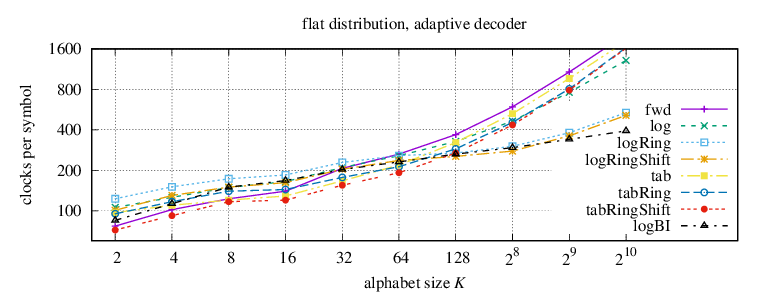}}
	\caption{\label{fig_clocksPerSymbolAdaptiveFlat}Adaptive mode comparison for uniform distribution: (a) encoder, (b) decoder}
\end{figure}

	%
	%
\Fig{fig_clocksPerSymbolAdaptiveFlat} shows the results for the encoder and decoder processes when applied to a flat distribution.
The computational load is now dominated by the adaptation of the cumulative counts.
In encoder, the conventional linear update ({\scshape updateSumCounts()}), which is used in combination with `fwd' and `log', seems to be preferable up to an alphabet size of $K=32$. 
The ring-buffer-based adaptation ({\scshape RCupdateModel()}) is not very competitive in this range. Albeit there are less accesses to the array of cumulative counts when using the ring-buffer technique, the overhead of updating the ring buffer itself seems to slow down the processing. Beyond $K=32$, the benefits of the binary-indexing method apply.

At decoder side, ring-buffer adaptation and shift operation in combination with the very fast table-based symbol determination make the proposed scheme `tabRingShift' the best choice for alphabets with $K<128$. For larger alphabets, the logarithmic symbol determination in combination with ring-buffer-based adaptation seems to be favourable. Probably, updating the table gets too costly for larger alphabets.
	%
\subsubsection{Geometric distribution}
\begin{figure}
    \captionsetup[subfloat]{labelfont=small,textfont=scriptsize}
	\centering
	\subfloat[]{\includegraphics[scale = 1.]{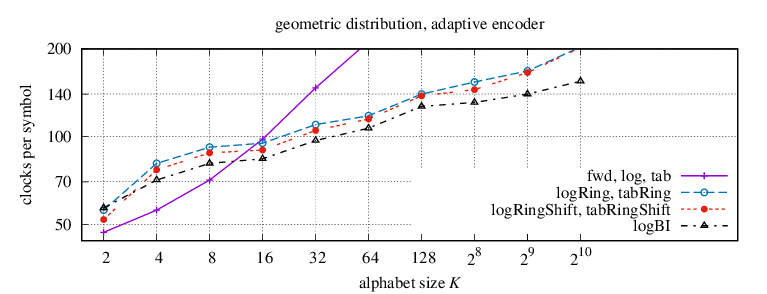}}\\[-1ex]

	\subfloat[]{\includegraphics[scale = 1.]{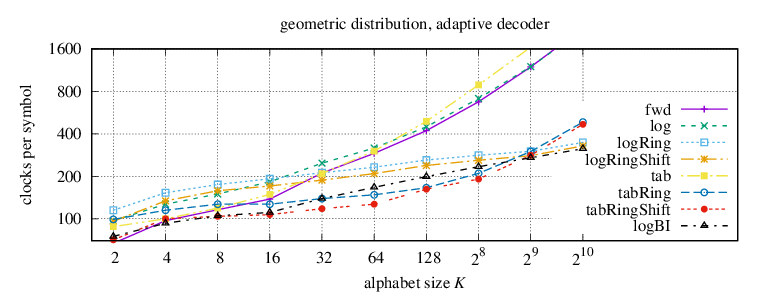}}	
	\caption{\label{fig_clocksPerSymbolAdaptiveGeometric}Adaptive mode comparison for geometric symbol distribution: (a) encoder, (b) decoder}
\end{figure}

	%
When processing geometric distributions, the behaviour is similar to uniform distribution, \Fig{fig_clocksPerSymbolAdaptiveGeometric}. 
Most algorithms can utilize the skewed distribution for faster processing. Only the linear update of cumulative counts has now a steeper curve, since on average more than $K/2$ cumulative counts have to be incremented per symbol.
	%
	%
\subsection{Influence of binary indexing}
	%
The remaining question to answer is, whether the advantages of table-based decoding and ring-buffer adaptation can be combined with the benefits of binary indexing.

Using a ring buffer provides the advantage that not all cumulative counts starting from the current symbol up to the end of the array have to be accessed. Instead, only the elements between the new and the old symbol (which is overwritten by the new) must be modified.
However, the binary index tree does not support such a range modification. The cumulative counts have first to be incremented according the new symbol and afterwards they have to be decremented according to the old symbol. The big-O complexity is still $\mathcal{O}(\log K)$, but the efforts have doubled.

A second problem arises if the table-based decoding is activated. In ring-buffer mode, the cumulative counts and the mapping table can be modified in a single loop. When using the binary indexing there is no such synergy effect and efforts of updating both arrays simply sum up.

\begin{figure}
    \captionsetup[subfloat]{labelfont=small,textfont=scriptsize}
	\centering
	\subfloat[]{\includegraphics[scale = 1.]{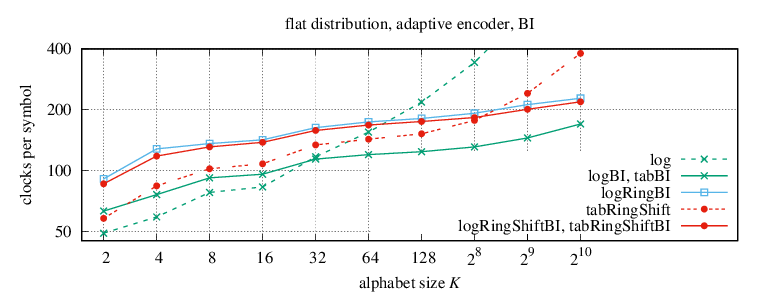}}\\[-1ex]

	\subfloat[]{\includegraphics[scale = 1.]{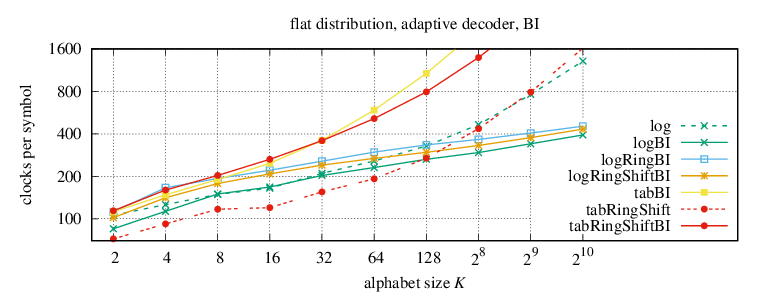}}	
	\caption{\label{fig_clocksPerSymbolAdaptiveFlatBI}Adaptive mode comparison for uniform symbol distribution when using binary indexing: (a) encoder, (b) decoder}
\end{figure}

\Figu{fig_clocksPerSymbolAdaptiveFlatBI} illustrates this set of problems for uniform distributions.
The curves for approaches applying the binary indexing have the same colours and point types as before, only the lines are drawn as solid lines.
For comparison, also the best results without binary indexing are shown.
The encoder significantly benefits from BI for $K\!\ge 32$, Fig.\ref{fig_clocksPerSymbolAdaptiveFlatBI}(a). For smaller alphabets the standard addressing is faster. The combination of BI with ring-buffer adaptation (`logRingBI') slows the processing down for reasons already explained above. The options `log' versus `tab' refer to the symbol identification in decoder and have no influence on the encoder.

The decoder plots, Figure \ref{fig_clocksPerSymbolAdaptiveFlatBI}(b), shows a slightly different picture. The proposed approach `tabRingShift' without BI performs best for $K\le 128$. The combination of BI and table-based decoding dramatically increase the adaptation costs making it impossible to benefit from their advantages.

\begin{figure}
    \captionsetup[subfloat]{labelfont=small,textfont=scriptsize}
	\centering
	\subfloat[]{\includegraphics[scale = 1.]{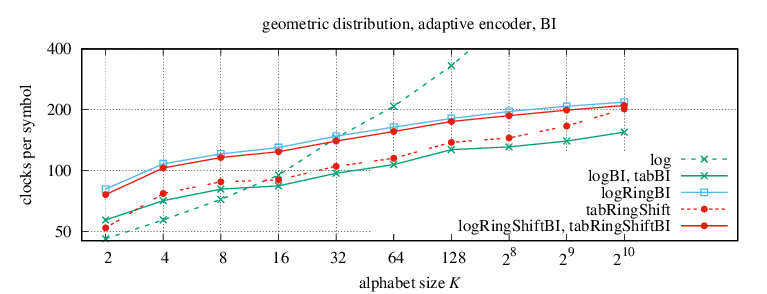}}\\[-1ex]
	
	\subfloat[]{\includegraphics[scale = 1.]{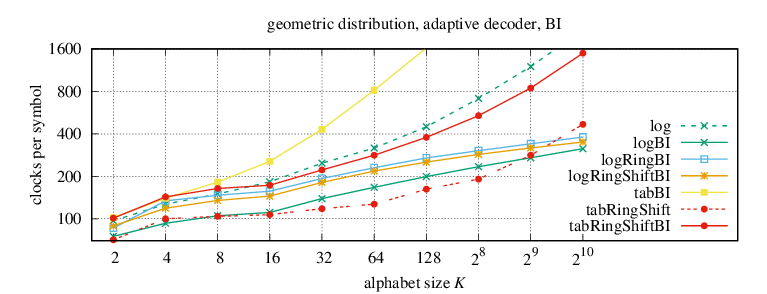}}
	\caption{\label{fig_clocksPerSymbolAdaptiveGeometricBI}Adaptive mode comparison for geometric symbol distribution when using binary indexing: (a) encoder, (b) decoder}
\end{figure}

The investigations based on the datasets with geometric distribution of symbols lead to similar results, whereas `tabRingShift' benefits most from this skewed distribution, \Figu{fig_clocksPerSymbolAdaptiveGeometricBI}.
	%
\subsection{Comparison of different settings}
	%
While in static coding the combination of table-based decoding, ring-buffer and substitution of a division with a bit-shift operation (tabRingShift) always performs fastest, in adaptive mode it depends of the alphabet size. \Tab{tab_recommendation} provides a rough overview of the best choice given a certain application mode and data type.
\begin{table*}
	\caption{\label{tab_recommendation}Recommended settings for different alphabet sizes. Labels are explained in Table \ref{tab_combinations}}
	\centering
	{\small\tabcolsep3pt
	\begin{tabular}{|c|c|c|c|c|c|c|}
	\hline
												& \multicolumn{2}{c|}{Encoder}  & \multicolumn{2}{c|}{Decoder} 									& \multicolumn{2}{c|}{Encoder + Decoder} 									\\
												& \multicolumn{2}{c|}{adaptive} & \multicolumn{2}{c|}{adaptive}	& \multicolumn{2}{c|}{adaptive}	 \\
	 	$K$									& flat 		& geom.								& flat & geom. 									& flat 					& geom. 									  \\
	\hline                                                                                  
	  $\sim 2\dots 12$		&	log 		&	log  								&	tabRingShift	&	tabRingShift		&	fwd						&	fwd			\\
	  $\sim 13\dots 32$		&	log 		& logBI	 							&	tabRingShift	&	tabRingShift		&	tabRingShift	&	tabRingShift			\\
	  $\sim 33 \dots 80$ 	&	logBI		&	logBI								& tabRingShift	&	tabRingShift		& tabRingShift	&	tabRingShift			\\
	  $\sim 81 \dots 500$	&	logBI		&	logBI 							& logBI					&	tabRingShift		& logBI					&	tabRingShift			\\
	 $\sim 501 \dots 1024$&	logBI		&	logBI 							& logBI					&	logBI						& logBI					&	logBI							\\
	\hline
	\end{tabular}
	}
\end{table*}
	%
On decoder side, the proposed scheme is the best option for alphabet sizes to about $K\!=\!80$ (uniform symbol distribution) or $K\!=\!500$ (geometric distribution).

On encoder side, there is no advantage of fast table-based processing.
 The adaptation of cumulated counts dominates the time-consumption here. Dependent on the symbol's distribution, the standard linear adaptation is favourable up to $K=32$ or $K=12$, respectively. Maintaining the ring buffer takes to much efforts in relation. For larger alphabets, the logarithmic complexity of the binary-indexing technique allows faster modifications.
This also can be observed for the decoder, where BI becomes an option for about $K\ge 81$.

\begin{figure}
	\centering
	\subfloat[]{\includegraphics[scale = 1.]{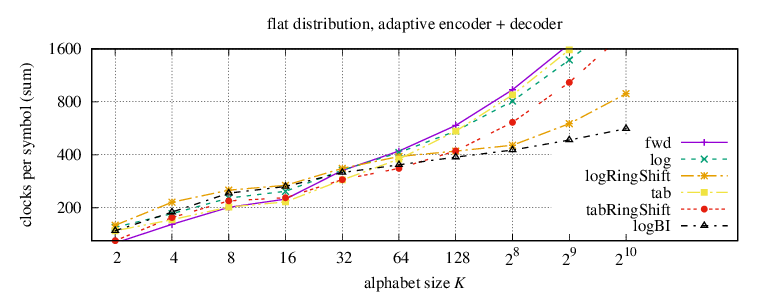}}\\[-1ex]
	
	\subfloat[]{\includegraphics[scale = 1.]{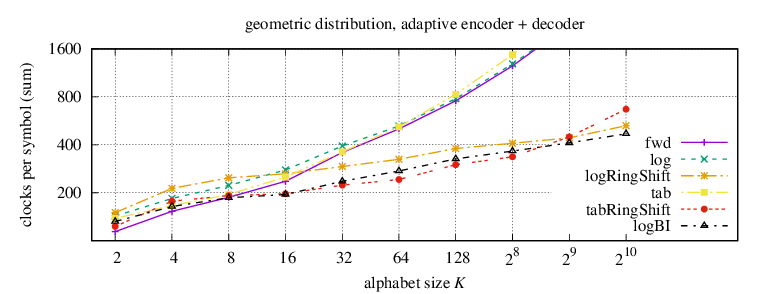}}
	\caption{\label{fig_clocksPerSymbolAdaptiveSum}Total processing times (encoder+decoder): (a) uniform distribution, (b) geometric distribution}
\end{figure}

The problem is, that encoder and decoder must apply the same adaptation strategy to be compatible: either both use the ring buffer or both do not. If the application demands either for fast encoding or decoding, Table \ref{tab_recommendation} can be used for guidance. If the overall time consumption is important, we have to investigate the sum of encoder and decoder times, \Figu{fig_clocksPerSymbolAdaptiveSum} and rightmost column in Table \ref{tab_recommendation}.
It can be seen that the proposed `tabRingShift' approach is a good compromise for alphabets with $K\le 128$ (flat) or $K\le 256$ (geometric) different symbols, respectively.
	%
	%
\subsection{Investigation of the ring-buffer approach}
	%
\subsubsection{Influence of the ring-buffer size on the bitrate}
\begin{figure}
		\centering
  \includegraphics[scale = 1.]{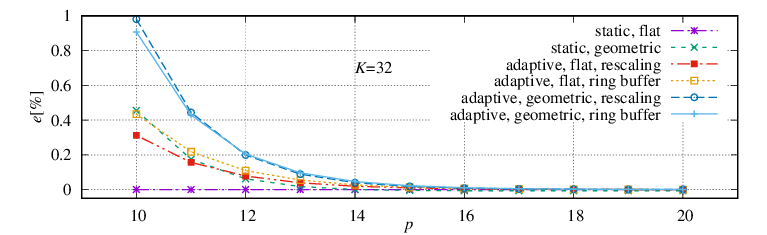}\\
	(a)
	
  \includegraphics[scale = 1.]{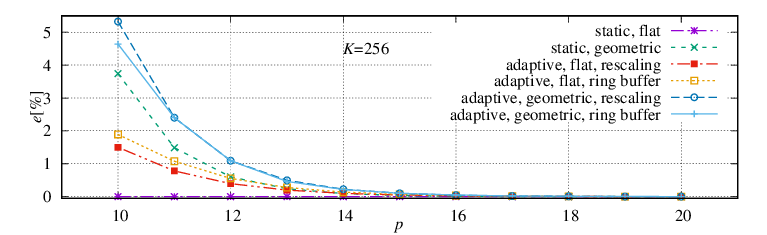}\\
	(b)
	\caption{\label{fig_bitrate}Accuracy of entropy coding in terms of difference between bitrate and signal entropy: $e = 100\cdot (R -H) / H$ [\%] dependent on the chosen total count $M=2^p$ and different alphabet sizes $K\in\{32,256\}$}
\end{figure}

	%
The size $M\!-\!K$ of the ring-buffer array influences two things: (i) the number of symbols to be processed before the algorithm can switch from division to bit-shift operation and (ii) how precise the cumulative counts $h_{\rm k}[\cdot]$ can represent the  cumulative probabilities $p_{\rm k}[\cdot]$. \Figu{fig_bitrate} shows the difference between the achieved bitrate (number of bits per encoded symbol) and the signal entropy $H$ for the ring-buffer approach in comparison to the standard approach with periodic rescaling of cumulative counts.
	%

In Figure \ref{fig_bitrate}(a), the datasets with $K\!=\!32$ different symbols have been exemplary chosen to investigate the performance.
The static mode can perfectly match the uniform distribution with an entropy of $H= 5.0$ bit/symbol as this distribution could be represented with counts of only $h[i]=1~\forall i$.
The more non-uniform a distribution is, the more precise the probabilities should be. The chosen geometric distribution has an entropy of $H= 2.978335$ bit/symbol. It can be seen that the achieved bitrate deviates from the entropy by about 0.1\% if less then $p=13$ bits are used for the entire interval.

For larger alphabet sizes it becomes more difficult to represent the distribution with a limited value of total counts, Figure \ref{fig_bitrate}(b).
When $M$ is fixed at $2^p$, the average number of available counts per symbol decreases as the alphabet grows. For $p = 12$, an average of 128 counts per symbol are available when $K = 32$, whereas when $K = 256$, only 16 are available. This results in coarser quantisation of the probabilities, which has a negative impact on the bitrate, particularly in the case of non-uniform distributions.

The investigations show that, for the same total count, the ring buffer approach offers roughly the same accuracy as rescaling the counts. With a uniform distribution and a very small buffer size, rescaling offers a slight advantage over the use of a ring buffer, as the former employs exponential decay and thus has a practically longer memory. When applied to the chosen geometric distribution, this difference disappears.
	%
\subsubsection{Influence of the ring-buffer size on the processing speed}
\begin{figure}
		\centering
  \includegraphics[scale = 1.]{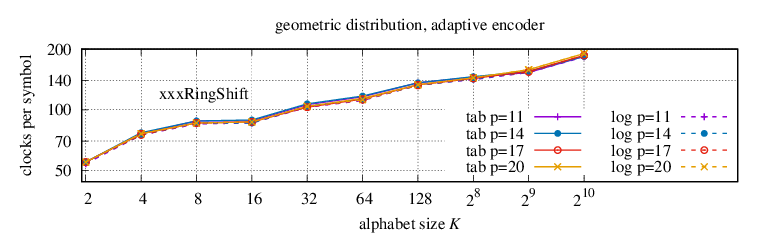}\\
	(a)
	
  \includegraphics[scale = 1.]{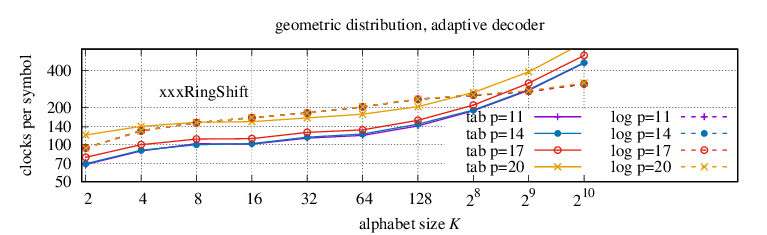}\\
	(b)
	\caption{\label{fig_speed}Processing time of (a) encoder and (b) decoder dependent on the chosen total count $M=2^p$}
\end{figure}

	%
\Fig{fig_speed} visualises how the processing speed changes depending on the chosen total count $M=2^p$ for configurations {\tt tabRingShift} and {\tt logRingShift}.
According to Fig. \ref{fig_speed}(a), the encoder speed appears to be independent on the ring-buffer size, although the switching from division to shift operation is delayed for larger sizes. There is no difference between  {\tt tab} and  {\tt log}, since these categories only apply to the symbol identification at the decoder. 
	
The decoding curves in Fig. \ref{fig_speed}(b) show no dependence when the logarithmic search is used in the decoder; however, they show a noticeable increase in processing time for very large values of $M$ when table-based symbol determination is used. It is possible that larger mapping tables do not fit into the CPU cache, and  accesses take longer (see Algorithm \ref{alg_RCupdateRingModel}, lines 10 and 13, and Algorithm \ref{alg_tableSearch}, lines 12--14).
	%
\subsubsection{Application to real data with varying statistics}
\begin{figure}
		\centering
  \includegraphics[scale = 1.]{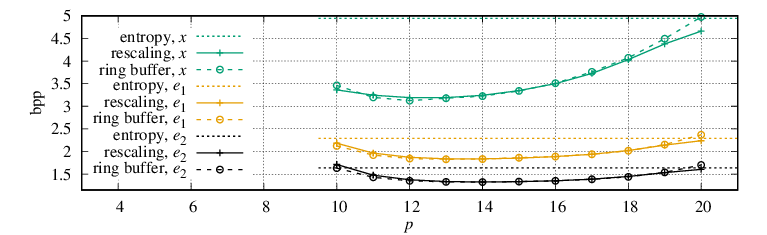}
	\caption{\label{fig_bitrateReal}Compression efficiency of ring-buffer and rescaling adaptation dependent on the chosen total count $M=2^p$ for test images $x$, $e_1$, and $e_2$}
\end{figure}

	%
In real data, symbol probabilities often vary depending on which part of the symbol sequence is in focus. 
This is the main application of adaptive coding.
Also here we investigate the compression efficiency of ring-buffer adaptation in comparison to conventional increment-and-rescale adaptation.
Three images have been chosen which were already used as test data in \cite{Str23}, see \cite{Dat25}. Image $x$ is an RGB image containing screen content, while images $e_1$ and $e_2$ are both prediction error images of $x$ created by different prediction methods. The images are stored as raw data in planar mode RRR..GGG..BBB.
\Figu{fig_bitrateReal} shows the resulting bitrates dependent on the chosen total count $M$.
The entropy $H(x)=4.942$~bit/pixel of image $x$ is drawn as a straight dotted line. Both, ring-buffer and rescaling adaptation, are able to compress the image with less bits per pixel (bpp) since the symbol distribution is drawn from a local context when using these adaptation techniques. The optimum (lowest bitrate) seems to be around $M=2^{12}$ for both approaches.

The two prediction error images have originally lower entropies ($H(e_1)=2.289$~bpp, $H(e_2)=1.638$~bpp) and adaptive coding including forgetting of old symbols facilitates saving at even lower bitrates. The distribution of the prediction error signals changes somewhat slower, therefore a longer memory ($p=14$) is beneficial. 
	%
\section{Summary}
\label{sec_Summary}
	%
Interval-based coding methods like conventional arithmetic coding, range coding, and range ANS require three time-consuming processing steps: the symbol search at the decoder, the core calculations at encoder and decoder, and, when operating in adaptive mode, the adaptation of cumulative counts at encoder and decoder.

We presented a new adaptation approach which can be used in conjunction with the very fast table-based symbol determination at decoder. It is based on storing the most recent symbols in a ring buffer. If the size of this buffer is appropriately chosen, the total sum of all counts can be kept constant equal to a power of two. This facilitates the substituting of a division with a bit-shift operation at both, encoder and decoder. 

The performance (speed and bitrate) of the proposed scheme has been investigated in comparison with other promising approaches by carrying out practical simulations. The investigations included different alphabet sizes $2 \le K \le 1024$, flat and geometric symbol distribution, and two modes of operation (static and adaptive).
	%
In adaptive compression mode, where updating the symbol statistics constitutes the major part of the workload, the proposed scheme is among the fastest approaches. The decoder processing time is typically shorter than that of other approaches. When the total time of encoder and decoder is considered, the proposed ring-buffer approach performs best for a wide range of alphabet sizes. Only when the number of distinct symbols becomes very large does accessing the cumulative counts via binary indexing outperform the proposed scheme.

The compression effectiveness is comparable to that of the conventional increment-and-rescale method. 
A drawback of the proposed scheme is that the advantage of replacing divisions by shift operations becomes significant only for sequences that are considerably longer than the chosen buffer size.
	%
	%
	%
	%
	%
\appendix
{
	%
\section{Derivations}
	%
\subsection{Number of accesses to array of cumulative counts}
	%
Given an array of $K$ elements, we are interested in the average number of accesses that are required to update the array content.
	%
\subsubsection{Assumed uniform distribution of symbols}\label{app_uniform}
	%
The conventional approach gets an index $i\in [0;K)$ as input and requires to access all elements from $i+1$ to $K$, thus the number of accesses is $x = K-i$. If $i$ follows a uniform distribution, the expectation value of $x$ is
\begin{align}
	E[X] &= \frac{1}{K}\tdot \sum_{i=0}^{K-1} K-i
	= \frac{1}{K}\tdot \left[K^2 -\sum_{i=0}^{K-1} i \right]  \nonumber\\
	&= \frac{1}{K}\tdot \left[K^2 - \frac{(K\!-\!1)K}{2} \right] 
	= K - \frac{K\!-\!1}{2}\approx \frac{K}{2}
			\;.
\end{align}
The proposed ring-buffer approach uses two indices $i\in [0;K)$ and $j\in [0;K)$ as input and requires to access all elements from $i$ to $j$, thus the number of accesses is $d = |j-i|$. 
There are exactly $2\cdot(K-d)$ pairs $(i,j)$.
If $i$ and $j$ follow a uniform distribution, the expectation value of $d$ over all possible $K^2$ pairs is
\begin{align}
	E[D] &= \frac{1}{K^2}\tdot \sum_{d=0}^{K-1} 2\tdot(K-d)\tdot d 
	= \frac{2}{K^2}\tdot \left[K\sum_{x=0}^{K-1}d - \sum_{d=0}^{K-1} d^2 \right]  \nonumber\\
	&= \frac{2}{K^2}\cdot \left[K\frac{(K-1)K}{2} - \frac{(K-1)K(2K-1)}{6}\right] \nonumber\\
	&= (K -1) - \frac{(K-1)(2K-1)}{3K} \\
	&=  - \frac{3K(K\!-\!1) - (K\!-\!1)(2K\!-\!1)}{3K} 
	=  \frac{K^2\! -\! 1}{3K} \approx \frac{K}{3}\nonumber
			\;.
\end{align}
	%
	%
\subsubsection{Assumed geometric distribution of symbols}\label{app_geometric}
	%
Symbols $i$ following an unbounded geometric distribution have a probability of $P(i\!=\!n)=(1-p)\cdot p^n$.  This unbounded (not truncated) distribution is used for simplifying the derivations and provides an upper bound of the average number of accesses.

In conventional approach, all elements from $i+1$ to $K$ must be modified. 
Assuming that symbols with highest index ($i=K-1$) have the highest probability, the average number of accesses $x$ can be approximated by the expectation value of the geometric distribution:
\begin{align}
		E[X] & 	\approx \frac{\displaystyle p}{\displaystyle 1-p}
			\;.
\end{align}
In the typical case however, symbols with low indices have the highest probability and the average number of accesses is about $K-E[X]$.

Using the ring-buffer approach, the distance between two indices $i$ and $j$  (= number of accesses) can be expressed as $d=|i-j| = i + j - 2\cdot \min(i,j)$.
So, the expected distance is
\begin{align}\label{eq_Ed}
	E[D] &= 2\tdot E[i] - 2\tdot E[\min(i,j)] 
			= \frac{2\tdot p}{1-p} - 2\tdot E[\min(i,j)]
			\;.
\end{align}
According to the tail-sum formula, the expectation value of the minimum can be expressed as
\begin{align}\label{eq_Eminij}
	E[\min(i,j)] = \sum_{n=1}^\infty P(\min(i,j)\ge n)
	\;.
\end{align}
It is further assumed that $i$ and $j$ are independent and identically distributed. Then the probability that $\min(i,j)$ is not smaller than $n$ is 
\begin{align}
	P(\min(i,j)\ge n)&=P(i\ge n \wedge j\ge n) 
										=P(i\ge n)\cdot P(j\ge n)				\nonumber\\
										&=P(i\ge n)^2=p^{2n}
	\;.
\end{align}
So, we can formulate (\ref{eq_Ed}) using (\ref{eq_Eminij}) as
\begin{align}
	E[D] &= \frac{2\tdot p}{1-p}  - 2\tdot \sum_{n=1}^\infty p^{2n} 
			 = \frac{2\tdot p}{1-p}  - 2\tdot\frac{p^2}{1-p^2} \nonumber\\
			&= \frac{2\tdot p\cdot (1+p)}{(1-p)(1+p)}  - 2\tdot\frac{p^2}{(1-p)(1+p)} 
			 = \frac{2p }{1-p^2}
			\;.
\end{align}
This average distance $E[D]$ or number of accesses depends on probability $p$, Equation (\ref{eq_geometricDistribution}), which itself depends on the parameter $k$, see (\ref{eq_kbasedOnK}). 
The actual complexity with respect to the investigated alphabet size $K$ is shown in \Tabl{tab_access}.
\begin{table*}
\caption{\label{tab_access}Geometrically distributed symbols: comparison of average number of accesses needed for updating the array of cumulative counts}
{\small\setlength{\tabcolsep}{3pt}
\hfil\begin{tabular}{|c|cc|ccc|cc|}
\hline
	 & \multicolumn{2}{c|}{parameter of}	& \multicolumn{3}{c|}{standard approach} & \multicolumn{2}{c|}{ring-buffer approach}\\
	 & \multicolumn{2}{c|}{distribution}	& \multicolumn{2}{c|}{best case}& worst case& 	&\\
	$K$ & $k$	& $p$							&$E[X] = \cfrac{\D p}{\D 1-p}$&  \multicolumn{1}{c|}{$\cfrac{\D E[X]}{\D K}\cdot K$} &  $\cfrac{\D K-E[X]}{\D K}\cdot K$ & $E[D]=\cfrac{\D 2\cdot p}{\D 1-p^2}$	& $\cfrac{\D E[D]}{\D K}\cdot K$	\\
	\hline
	\hline
	2		&		0	& 0.5 						&	1				&	$1/2\cdot K$		&$1/2\cdot K$			&	4/3					&		$2/3\cdot K$	\\	
	4		&		0	& 0.5 						&	1				&	$1/4\cdot K$		&$3/4\cdot K$			&	4/3					&		$1/3\cdot K$			\\	
	8		&		0	& 0.5 						&	1				&	$1/8\cdot K$		&$7/8\cdot K$			&	4/3					&		$1/6\cdot K$			\\	
	16	&		0	& 0.5 						&	1				&	$1/16\cdot K$		&$15/16\cdot K$			&	4/3					&		$1/12\cdot K$		\\	
	32	&		1	& $1/\sqrt{2}$ 		&	2.4142	&	$0.0754\cdot K$&	$0.9246\cdot K$&	2.8284			&		$0.0884\cdot K$		\\	
	64	&		2	& $1/\sqrt[4]{2}$ &	5.2852	&	$0.0826\cdot K$&	$0.9174\cdot K$&	5.7420			&		$0.0897\cdot K$	\\	
	128	&		3	& $1/\sqrt[8]{2}$ &	11.0488	&	$0.0863\cdot K$&	$0.9137\cdot K$&	11.527			&		$0.0901\cdot K$	\\	
	256	&		4	& $1/\sqrt[16]{2}$ &22.5867	&	$0.0882\cdot K$&	$0.9118\cdot K$&	23.076			&		$0.0901\cdot K$	\\	
	512	&		5	& $1/\sqrt[32]{2}$ &45.6680	&	$0.0892\cdot K$&	$0.9108\cdot K$&	46.163			&		$0.0902\cdot K$	\\	
1024	&		6	& $1/\sqrt[64]{2}$ &91.8334	&	$0.0900\cdot K$&	$0.9103\cdot K$&	92.331			&		$0.0902\cdot K$	\\
\hline
\end{tabular}
}
\end{table*}

For large $K$, these values provide a good approximation of the complexity. For $K\le 8$, the actual average number of accesses is lower, since $d$ cannot be arbitrarily high for the truncated distribution used.
} 

		%
	%
\ifdraft
\bibliographystyle{alphaTS}
\fi
\bibliography{literature} 

\begin{thebibliography}{Ram20}
\providecommand{\url}[1]{\texttt{#1}}
\providecommand{\urlprefix}{URL }
\expandafter\ifx\csname urlstyle\endcsname\relax
  \providecommand{\doi}[1]{doi:\discretionary{}{}{}#1}\else
  \providecommand{\doi}{doi:\discretionary{}{}{}\begingroup
  \urlstyle{rm}\Url}\fi
\providecommand{\selectlanguage}[1]{\relax}

\bibitem[AL16]{Aul16}
Aul\'{i}-Llin\`{a}s, F.; Enfedaque, P.; Moure, J.C.; Sanchez, V.: Bitplane
  Image Coding With Parallel Coefficient Processing.
\newblock \emph{IEEE Transactions on Image Processing}, 2016.
\newblock Vol.~25, No.~1, 209--219

\bibitem[Bel13]{Bel13}
Belyaev, E.; Turlikov, A.; Egiazarian, K.; Gabbouj, M.: An Efficient Adaptive
  Binary Arithmetic Coder With Low Memory Requirement.
\newblock \emph{IEEE Journal of Selected Topics in Signal Processing}, 2013.
\newblock Vol.~7, No.~6, 1053--1061

\bibitem[Bel17]{Belyaev2017}
Belyaev, E.; Forchhammer, S.; Liu, K.: An Adaptive Multialphabet Arithmetic
  Coding Based on Generalized Virtual Sliding Window.
\newblock \emph{IEEE Signal Processing Letters}, 2017.
\newblock Vol.~24, No.~7, 1034--1038

\bibitem[Bos21]{Bos21}
Bossen, F.; S{\"u}hring, K.; Wieckowski, A.; Liu, S.: {VVC} Complexity and
  Software Implementation Analysis.
\newblock \emph{IEEE Transactions on Circuits and Systems for Video
  Technology}, 2021.
\newblock Vol.~31, No.~10, 3765--3778

\bibitem[Cai22]{Cai22}
Cai, Y.; Fan, Y.; Huang, L.; Zeng, X.; Yin, H.; Zeng, B.: A Fast {CABAC}
  Hardware Design for Accelerating the Rate Estimation in {HEVC}.
\newblock \emph{IEEE Transactions on Circuits and Systems for Video
  Technology}, 2022.
\newblock Vol.~32, No.~4, 2385--2395

\bibitem[Che15]{Che15}
Chen, Y.H.; Sze, V.: A Deeply Pipelined {CABAC} Decoder for {HEVC} Supporting
  Level 6.2 High-Tier Applications.
\newblock \emph{IEEE Trans. on Circuits and Systems for Video Technology},
  2015.
\newblock Vol.~25, No.~5, 856--868

\bibitem[Che18]{Chen18}
Chen, Y.; Murherjee, D.; Han, J.; Grange, A.; Xu, Y.; Liu, Z.; Parker, S.;
  Chen, C.; Su, H.; Joshi, U.; Chiang, C.H.; Wang, Y.; Wilkins, P.; Bankoski,
  J.; Trudeau, L.; Egge, N.; Valin, J.M.; Davies, T.; Midtskogen, S.; Norkin,
  A.; de~Rivaz, P.: An Overview of Core Coding Tools in the {AV1} Video Codec.
\newblock In \emph{2018 Picture Coding Symposium (PCS)}. 2018, 41--45

\bibitem[Che20]{ChS20}
Chen, M.; Su, H.; Deng, S.; Xu, Y.: Machine Learning Based Symbol Probability
  Distribution Prediction For Entropy Coding In {AV1}.
\newblock In \emph{2020 IEEE International Conference on Image Processing
  (ICIP)}. 2020, 3374--3378

\bibitem[Cho23]{Chou23}
Chou, Y.G.; Chen, J.J.: {H.266/VVC} Time Complexity Reduction by Learned Models
  and Image Statistical Features.
\newblock In \emph{2023 IEEE International Conference on Visual Communications
  and Image Processing ({VCIP})}. 2023, 1--5

\bibitem[Col15]{Collet2015}
Collet, Y.: Zstandard - fast real-time compression algorithm, 2015

\bibitem[Dos13]{Doshi2013}
Doshi, J.; Gandhi, S.: Enhanced arithmetic coding using total frequency in
  power of 2 \& processing multi-bits at a time.
\newblock In \emph{2013 Sixth International Conference on Contemporary
  Computing (IC3)}. 2013, 1--6

\bibitem[Dud14]{Dud14}
Duda, J.: Asymmetric numeral systems: entropy coding combining speed of
  {Huffman} coding with compression rate of arithmetic coding, 2014

\bibitem[Dud15]{Dud15}
Duda, J.; Tahboub, K.; Gadgil, N.J.; Delp, E.J.: The use of asymmetric numeral
  systems as an accurate replacement for Huffman coding.
\newblock In \emph{2015 Picture Coding Symposium (PCS)}. 2015, 65--69

\bibitem[Fen94]{Fen94}
Fenwick, P.: A new data structure for cumulative probability tables.
\newblock \emph{Software: Practice and Experience}, March 1994.
\newblock Vol.~24, No.~3, 327--336

\bibitem[Gal78]{Gallager1978}
Gallager, R.: Variations on a theme by Huffman.
\newblock \emph{IEEE Transactions on Information Theory}, 1978.
\newblock Vol.~24, No.~6, 668--674

\bibitem[Gha91]{Ghanbari1991}
Ghanbari, M.: Arithmetic coding with limited past history.
\newblock \emph{Electronics Letters}, 1991.
\newblock Vol.~27, No.~13

\bibitem[Gon23]{Gon23}
Goncalves, P.; Correa, G.; Agostini, L.; Porto, M.: Learning-based bypass zone
  search algorithm for fast motion estimation.
\newblock \emph{Multimedia Tools and Applications}, 2023.
\newblock Vol.~82, No.~3, 3535--3560

\bibitem[Han22]{Han22}
Han, J.; Xu, Y.: Probability Model Estimation for M-Ary Random Variables.
\newblock In \emph{2022 IEEE International Conference on Image Processing
  (ICIP)}. 2022, 3021--3025

\bibitem[How92]{Howard1992}
Howard, P.G.; Vitter, J.S.: \emph{Practical Implementations of Arithmetic
  Coding}.
\newblock Springer US, Boston, MA, 1992.
\newblock ISBN 978-1-4615-3596-6, 85--112

\bibitem[Hsi22]{Hsi22}
Hsieh, P.A.; Wu, J.L.: A Review of the Asymmetric Numeral System and Its
  Applications to Digital Images.
\newblock \emph{Entropy}, 2022.
\newblock Vol.~24, No.~3

\bibitem[Jia19]{Jia19}
Jia, L.; Tsui, C.Y.; Au, O.C.; Jia, K.: A New Rate-Complexity-Distortion Model
  for Fast Motion Estimation Algorithm in {HEVC}.
\newblock \emph{IEEE Transactions on Multimedia}, 2019.
\newblock Vol.~21, No.~4, 835--850

\bibitem[Kim25]{Kim25}
Kim, S.; Jo, B.; Beack, S.; Lee, T.; Jun, D.: Dynamic Bit-Plane Arithmetic
  Coding Method for Quantized Spectral Coefficients in {USAC}.
\newblock \emph{IEEE Journal of Selected Topics in Signal Processing}, 2025,
  1--10

\bibitem[Kri25]{Kri25}
Krishnan, M.P.; Zhao, L.; Dong, Z.; Liu, T.; Kuang, W.; Tang, M.; Liu, S.:
  Enhanced Frame Context Initialization for Video Coding Beyond {AV1}.
\newblock In \emph{2025 IEEE International Conference on Image Processing
  (ICIP)}. 2025, 1594--1599

\bibitem[Liu22]{Liu22}
Liu, Y.; Abdoli, M.; Guionnet, T.; Guillemot, C.; Roumy, A.: Statistical
  Analysis of Inter Coding in {VVC} Test Model ({VTM}).
\newblock In \emph{2022 IEEE Int. Conference on Image Processing ({ICIP})}.
  2022, 3456--3459

\bibitem[Liu24]{Liu24}
Liu, S.; Cui, S.; Li, T.; Liu, H.; Yang, Q.; Yang, H.: Fast {CU} partition
  algorithm based on swin-transformer for depth intra coding in {3D-HEVC}.
\newblock \emph{Multimedia Tools and Applications}, 2024

\bibitem[Man05]{Man05}
Man, H.; Docef, A.; Kossentini, F.: Performance Analysis of the {JPEG} 2000
  Image Coding Standard.
\newblock \emph{Multimedia Tools and Applications}, 2005.
\newblock Vol.~26, No.~1, 27--57

\bibitem[Mar79]{Mar79}
Martin, G.N.: Range encoding: An algorithm for removing redundancy from a
  digitized message.
\newblock In \emph{{\em Video \& Data Recording Conference}}, Vol. 245.
  Southampton, UK, 1979

\bibitem[Mar03]{Mar03}
Marpe, D.; Schwarz, H.; Wiegand, T.: Context-based adaptive binary arithmetic
  coding in the {H.264/AVC} video compression standard.
\newblock \emph{IEEE Transactions on Circuits and Systems for Video
  Technology}, 2003.
\newblock Vol.~13, No.~7, 620--636

\bibitem[Mer21]{Mer21}
Mercat, A.; M{\"a}kinen, A.; Sainio, J.; Lemmetti, A.; Viitanen, M.; Vanne, J.:
  Comparative Rate-Distortion-Complexity Analysis of {VVC} and {HEVC} Video
  Codecs.
\newblock \emph{IEEE Access}, 2021.
\newblock Vol.~9, 67813--67828

\bibitem[Mic]{Mic24}
Microsoft: \_\_rdtsc.
\newblock Available:
  \url{https://learn.microsoft.com/en-us/cpp/intrinsics/rdtsc?view=msvc-170}.
\newblock Accessed 23 April 2025

\bibitem[Mof99]{Mof99}
Moffat, A.: An improved data structure for cumulative probability tables.
\newblock \emph{Software: Practice and Experience}, 1999.
\newblock Vol.~29, No.~7, 647--659

\bibitem[Mof20]{Mof20}
Moffat, A.; Petri, M.: Large-Alphabet Semi-Static Entropy Coding Via Asymmetric
  Numeral Systems.
\newblock \emph{ACM Trans. Inf. Syst.}, Jul. 2020.
\newblock Vol.~38, No.~4

\bibitem[Niu23]{Niu23}
Niu, W.; Huang, X.; Yin, H.; Lu, Y.; Zhou, Y.; Yan, C.: Fast all zero block
  detection algorithm for versatile video coding.
\newblock \emph{Multimedia Tools and Applications}, 2023.
\newblock Vol.~82, No.~22, 33693 -- 33718

\bibitem[Och24]{Och24}
Och, H.; Uddehal, S.R.; Strutz, T.; Kaup, A.: Enhanced Color Palette Modeling
  For Lossless Screen Content Compression.
\newblock In \emph{ICASSP 2024 - 2024 IEEE International Conference on
  Acoustics, Speech and Signal Processing (ICASSP)}. 2024, 3670--3674

\bibitem[Pak20]{Pak20}
Pakdaman, F.; Adelimanesh, M.A.; Gabbouj, M.; Hashemi, M.R.: Complexity
  Analysis Of Next-Generation {VVC} Encoding And Decoding.
\newblock In \emph{2020 IEEE Int. Conf. on Image Proc. (ICIP)}. 2020,
  3134--3138

\bibitem[Ram20]{Ram20}
Ramos, F.L.L.; Saggiorato, A.V.P.; Zatt, B.; Porto, M.; Bampi, S.: Residual
  Syntax Elements Analysis and Design Targeting High-Throughput {HEVC CABAC}.
\newblock \emph{IEEE Transactions on Circuits and Systems I: Regular Papers},
  2020.
\newblock Vol.~67, No.~2, 475--488

\bibitem[Ris79]{Ris79}
Rissanen, J.; Langdon, G.: Arithmetic Coding.
\newblock \emph{IBM Journal of Research and Development}, March 1979.
\newblock Vol.~23, 149 -- 162

\bibitem[Rya99]{Ryabko1999}
Ryabko, B.; Fionov, A.: Fast and Space-Efficient Adaptive Arithmetic Coding.
\newblock In M.~Walker, editor, \emph{Cryptography and Coding}. Springer Berlin
  Heidelberg, Berlin, Heidelberg, 1999.
\newblock ISBN 978-3-540-46665-9, 270--279

\bibitem[Rya03]{Ryabko2003}
Ryabko, B.; Rissanen, J.: Fast adaptive arithmetic code for large alphabet
  sources with asymmetrical distributions.
\newblock \emph{IEEE Communications Letters}, 2003.
\newblock Vol.~7, No.~1, 33--35

\bibitem[Sai04]{Sai04}
Said, A.: Comparative analysis of arithmetic coding computational complexity.
\newblock In \emph{Data Compression Conference, 2004. Proceedings. DCC 2004}.
  2004, 562.
\newblock Also: techn. report at {HP Laboratories Palo Alto, HPL-2004-75}

\bibitem[Sal20]{Sal20}
Saldanha, M.; Sanchez, G.; Marcon, C.; Agostini, L.: Complexity Analysis Of
  {VVC} Intra Coding.
\newblock In \emph{2020 IEEE International Conference on Image Processing
  ({ICIP})}. 2020, 3119--3123

\bibitem[Sch98]{Schindler1998}
Schindler, M.: Range Coder.
\newblock Available: \url{www.compressconsult.com/rangecoder/}, 1998.
\newblock Accessed 04 November 2026

\bibitem[Sha23]{Sha23}
Shang, X.; Li, G.; Zhao, X.; Han, H.; Zuo, Y.: Fast {CU} size decision
  algorithm for {VVC} intra coding.
\newblock \emph{Multimedia Tools and Applications}, 2023.
\newblock Vol.~82, No.~18, 28301--28322

\bibitem[She23]{She23}
Sheng, Q.; Zhu, H.; Sheng, H.; Huang, X.; Jiang, J.; Lai, C.: High efficiency
  lossless image recompression algorithm with asymmetric numeral systems for
  real-time mobile application.
\newblock \emph{Journal of Real-Time Image Processing}, 2023.
\newblock Vol.~20, No.~5, 90

\bibitem[Str20]{Str19}
Strutz, T.; M\"{o}ller, P.: Screen Content Compression Based on Enhanced Soft
  Context Formation.
\newblock \emph{IEEE Transactions on Multimedia}, 2020.
\newblock Vol.~22, No.~5, 1126--1138

\bibitem[Str23]{Str23}
Strutz, T.: Rescaling of Symbol Counts for Adaptive r{ANS} Coding.
\newblock In \emph{31st European Signal Processing Conference (EUSIPCO)}.
  Helsinki, Finland, 2023, 585 -- 589

\bibitem[Str25a]{Dat25}
Strutz, T.: Data files for testing entropy coding software, 2025.
\newblock \doi{10.21227/40cv-tz15}

\bibitem[Str25b]{Stru25}
Strutz, T.; Schreiber, N.: Investigations on Algorithm Selection for
  Interval-Based Coding Methods.
\newblock \emph{Multimedia Tools and Applications}, July 2025

\bibitem[Sze12]{Sze12}
Sze, V.; Budagavi, M.: High Throughput {CABAC} Entropy Coding in {HEVC}.
\newblock \emph{IEEE Transactions on Circuits and Systems for Video
  Technology}, 2012.
\newblock Vol.~22, No.~12, 1778--1791

\bibitem[Wit87]{Witten1987}
Witten, I.; Neal, R.; Cleary, J.: Arithmetic Coding for data compression.
\newblock \emph{Communications of the ACM}, Juni 1987.
\newblock Vol.~30, 520 -- 540

\bibitem[Zha19]{Zhan19}
Zhang, Y.; Lu, C.: A Highly Parallel Hardware Architecture of Table-Based
  {CABAC} Bit Rate Estimator in an {HEVC} Intra Encoder.
\newblock \emph{IEEE Transactions on Circuits and Systems for Video
  Technology}, 2019.
\newblock Vol.~29, No.~5, 1544--1558

\bibitem[Zst26]{Zstandard}
Zstandard, 2026

\end{thebibliography}



\end{document}